\documentclass[lettersize,journal]{IEEEtran}
\usepackage{amsmath,amsfonts}
\usepackage{algorithmic}
\usepackage{algorithm}
\usepackage{array}
\usepackage[caption=false,font=normalsize,labelfont=sf,textfont=sf]{subfig}
\usepackage{textcomp}
\usepackage{stfloats}
\usepackage{url}
\usepackage{verbatim}
\usepackage{graphicx}
\usepackage{subfig}
\usepackage{cite}

\usepackage[colorlinks,
            linkcolor=black,
            anchorcolor=black,
            citecolor=black,
            urlcolor=black 
            ]{hyperref}
\usepackage{booktabs}
\usepackage{indentfirst}              
\usepackage{enumitem}
\setlist[itemize]{  
    leftmargin=2em,       
    labelsep=0.3em,       
    labelwidth=1.2em      
}

\usepackage{fancyvrb}
\usepackage{fvextra}

\begin{document}

\title{Visionary Co-Driver: Enhancing Driver Perception of Potential Risks with LLM and HUD}
\author{
	\IEEEauthorblockN{
		Wei Xiang\IEEEauthorrefmark{1}, 
		Ziyue Lei\IEEEauthorrefmark{1}, 
		Jie Wang\IEEEauthorrefmark{2}, 
        Yingying Huang\IEEEauthorrefmark{1},
		Qi Zheng\IEEEauthorrefmark{3}, 
		Tianyi Zhang\IEEEauthorrefmark{4},
            An Zhao\IEEEauthorrefmark{4},
		and Lingyun Sun\IEEEauthorrefmark{1}} 
    \IEEEauthorblockA{\IEEEauthorrefmark{1}International Design Institute, Zhejiang University, Hangzhou, China}
    
    \IEEEauthorblockA{\IEEEauthorrefmark{2}GRASP Lab, University of Pennsylvania, PA, USA}
    
    \IEEEauthorblockA{\IEEEauthorrefmark{3}The Hong Kong University of Science and Technology (Guangzhou), Guangzhou, China}
    
    \IEEEauthorblockA{\IEEEauthorrefmark{4}Chu Kochen Honors College, Zhejiang University, Hangzhou, China}
}

\markboth{Journal of \LaTeX\ Class Files,~Vol.~14, No.~8, August~2021}%
{Shell \MakeLowercase{\textit{et al.}}: A Sample Article Using IEEEtran.cls for IEEE Journals}


\maketitle

\begin{abstract}
Drivers' perception of risky situations has always been a challenge in driving. 
Existing risk-detection methods excel at identifying collisions but face challenges in assessing the behavior of road users in non-collision situations. 
This paper introduces Visionary Co-Driver, a system that leverages large language models (LLMs) to identify non-collision roadside risks and alert drivers based on their eye movements. 
Specifically, the system combines video processing algorithms and LLMs to identify potentially risky road users. 
These risks are dynamically indicated on an adaptive heads-up display interface to enhance drivers' attention. 
A user study with 41 drivers confirms that Visionary Co-Driver improves drivers' risk perception and supports their recognition of roadside risks.
\end{abstract}


\section{Introduction}
\IEEEPARstart{D}{rivers'} risk perception is a crucial factor of safe and efficient driving.
As driving automation becomes prevalent, driver's role transfers from operation to monitoring and supervision.
This shift leads to distractions, where drivers may not be fully engaged in driving \cite{HazardReview}.
Therefore, supporting drivers for comprehensive risk perception is essential \cite{tran2018human}.

Driving risk measures the possibility of collision and damage \cite{ieeeRiskPerception}.
Time-to-Collision (TTC) \cite{ttc} and Time Exposed Time-to-collision \cite{tet} are two factors that assess collision probability where the paths of road users may intersect with the vehicle's. Recently proposed scene-generalized models, such as Driver Risk Field \cite{drf} and Safety Field \cite{sf, RFNature}, use ``Field Strength'' to model the possibilities. However, the equally significant risks posed by roadside road users are frequently neglected. The paths of roadside road users would not intersect with the vehicle’s.
For example, pedestrians and non-motorized users suddenly swerve or rush onto the road. 
Because of their uncertain movement, there is no unified assessment on the risk of non-motorized vehicles and pedestrians \cite{ReviewPedBehave, HazardReview}.
This is acknowledged as a weakness of existing algorithms \cite{CoreChallengeAV}. 

Large language models (LLMs) exhibit exceptional common sense reasoning ability in diverse scenarios, making them promising tools for enhancing risk perception in non-collision situations \cite{wang2023surveyAgent, DriveLikeaHuman}. 
However, concerns regarding their reliability in driving, due to hallucinations and uncertainties, necessitate cautious integration into driver assistance systems.
This study proposes Visionary Co-Driver (VCD), a system that reminds drivers of neglected risks from roadside road users in non-collision situations.
Specifically, this paper employs LLMs to analyze road scene information and identify potential risks. Then, it use a HUD integrated with eye tracking to offer adaptive risk warnings and guidance.
A study involving 41 drivers evaluated the feasibility and effectiveness of our system.

To encapsulate, the contributions of this paper include:
\begin{itemize}

    \item \textbf{Using LLMs to assist risk recognition}: Our framework utilizes LLMs as a co-driver to analyze and alert potential risks. This framework can be extended to provide intelligent feedback for a broader range of cognitive support tasks in intelligent transportation systems.
    \item \textbf{Enhancing risk perception with HUD Interface}: We introduce a HUD interface, equipped with eye-tracking technology to dynamically adjust risk warnings based on the driver's gaze, providing a seamless integration with the LLM-based risk analysis, thereby enhancing overall driver awareness and safety.
    \item \textbf{Experimental Validation}: A user study with 41 drivers serves as empirical evidence to demonstrate the feasibility and effectiveness of VCD. 
\end{itemize}

\section{Literature Review}
\subsection{Challenges of Risk Perception in Driving}
\begin{figure}[!h]
    \centering
    \includegraphics[width=1\linewidth]{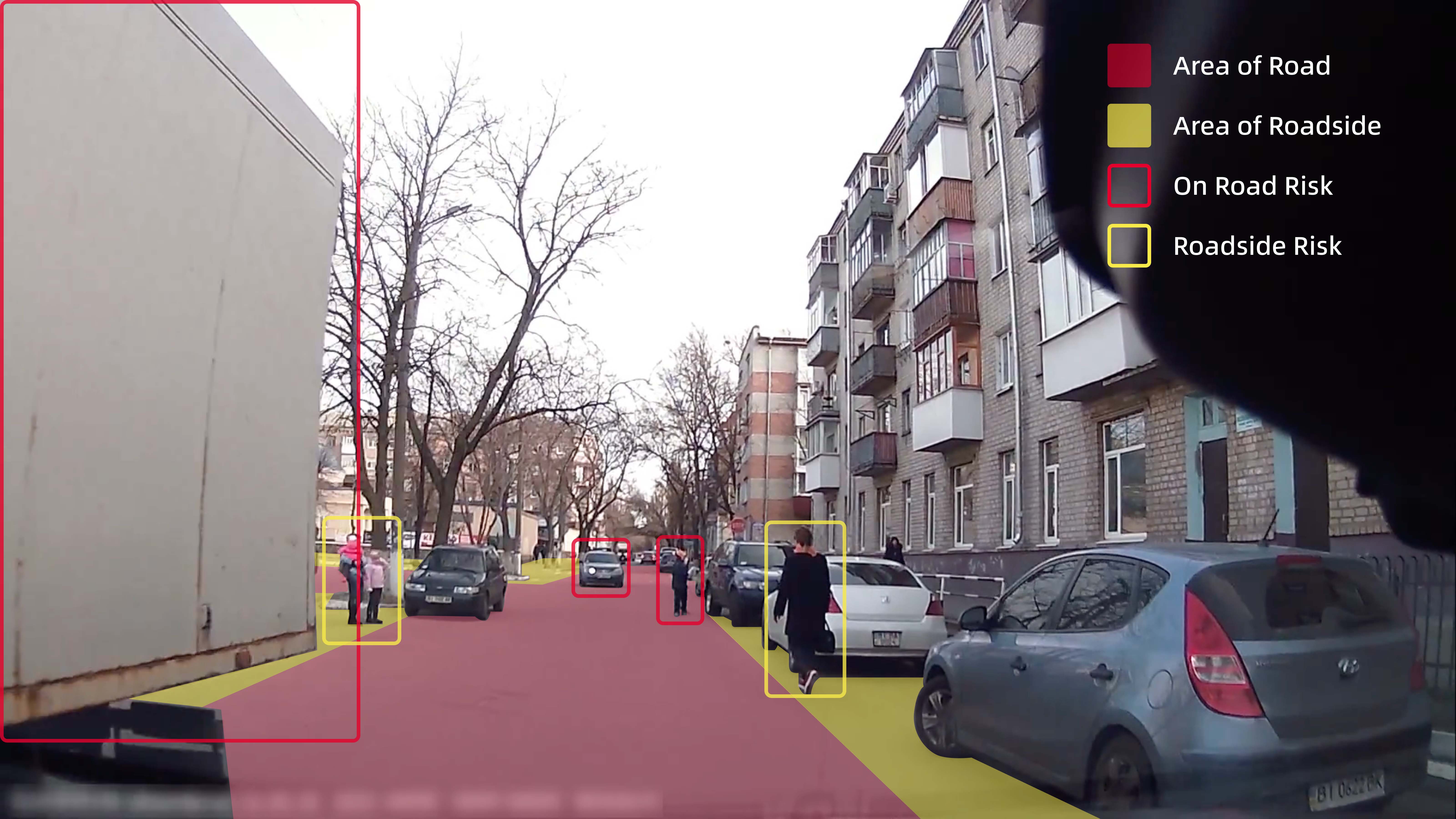}
    \caption{A typical road scene driver perceives in a cross road, with color-masked separation of ``on road'' and ``roadside'' area. On the left and far front, a truck, a car and a standing pedestrian is labeled with ``On Road Risk''. On the left and right, several standing pedestrian and a walking pedestrian is label with ``Roadside Risk''.}
    \label{fig:risk classification}
\end{figure}

Drivers need not only to detect risks on the road, but also risks on the roadside, such as sidewalks and lawns, as shown in Figure \ref{fig:risk classification}.
The behaviour of roadside pedestrians is fast changing and hard to predict.
They might cross the road even if there is no crosswalk, leaving drivers limited time to avoid crashes.
On complex urban roads, drivers may face difficulties in recognizing the crossing intentions of all road users.

Also, driving automation is negatively correlated with drivers' attention to the road ahead \cite{Automated_attention}.
Automation may reduce the physical demands on drivers while also increasing the potential for distraction. 
This can impede the ongoing vigilance and the capacity for rapid re-engagement with the driving task, both of which are essential for the effective management of risk in limited automated driving.

\subsection{Supporting Risk Perception with LLMs}
The stochastic nature of road user intentions poses challenges for accurately prediction \cite{ReviewPedBehave, CoreChallengeAV}.
Common risky behaviors include crossing the road, sudden stops, smartphone usage, and unexpected appearances of children or pets into the road. 
Existing road user intention estimation algorithms often struggle to adapt to the unpredictability and complexity of real-world driving environments \cite{DriveLikeaHuman}. 
To address this, various methods have been proposed, such as introducing more features of pedestrian movement patterns, multi-task sequence-to-sequence architectures, multiple model Kalman filtering and Long Short-Term Memory models to estimate pose and obtain pedestrian trajectories via cues from both pedestrians and vehicles \cite{ReviewPedBehave, huang2023learning,IntentionML,PoseRNN, LSTM_PedTrajectory}. 
These advancements show promise for comprehensive pedestrian intention estimation and contribute to improving the performance of prediction models. 
Still, it is challenging to handle corner cases using software simulation testing \cite{ChallengeAV_Test}.

Recently, the combination of LLMs and intelligent driving systems has become a heated discussion. This attention comes from LLMs' outstanding ability in comprehensive common-sense reasoning across various scenarios.
Using natural languages for task reasoning enables flexible, semantic-rich access and greater explain-ability \cite{wang2023surveyAgent, DriveLikeaHuman}.
Therefore, LLMs may assist driving in terms of corner case identification, human-like reasoning and autonomous decision-making \cite{stanford2023semantic, DriveLikeaHuman, ye2023largeDecision}.

While the idea of leveraging LLMs to analyze driving videos is intuitive and straightforward, video understanding using LLMs is still a problem. 
Firstly, the narrow conversational windows of LLMs restrict their reasoning capabilities \cite{wang2023surveyAgent}. This implies that while LLMs excel at processing textual information, they may struggle with the temporal expanse and complex contextual understanding required for video data, especially in scenarios that demand long-duration tracking and intricate spatial awareness. Secondly, LLMs performance in reasoning over non-natural language inputs is sub-optimal \cite{iNLP}, which is a critical issue considering that driving videos encompass not only visual elements but also complex temporal sequences and spatial relationships that require sophisticated understanding and inference.

Current approaches, such as the VideoLLM framework introduced by Chen et al., which converts video sequences into a unified token sequence \cite{chen2023videollm}, and the X-LLM, treating vision-to-text as a translation task, attempt to address this challenge from a pre-training perspective \cite{chen2023xllm}.
Additionally, Peng's novel prompt design method aims to enhance the extraction of semantic information \cite{peng2023prompt}, and the Stanford Autonomous Systems Lab has applied an object detection method at a 2Hz downsample rate \cite{stanford2023semantic}. There are also other efforts, like utilizing Visual Question Answering (VQA) models and Large Vision Language Models (LVLM) to directly extract information from videos \cite{otter, videogpt}. 
These approaches might support our exploration.

\subsection{HUD for Risk Warning}
HUD has been already widely used and shown remarkable effects \cite{distractive-or-supportive}. 
Existing HUDs are primarily designed to display immediate and concrete risks on road, such as proximity warnings triggered by distance sensors\cite{impact-of-position, ia-in-HUD}. Therefore, they drag users' attention to make sure these information are fully noticed. However, this design is overloading where the broad detection of roadside risks increases the amount of information displayed on the HUD interface.
Furthermore, roadside risks are uncertain and long-term, making HUD inappropriate to occupy an extended long duration of attention.

Integrating HUD with eye tracking could be a solution.
Although eye-tracking data has been widely studied for driver state monitoring \cite{InattentionMonitoringSystem}, few researchers have attempted to give timely feedback to drivers using eye-tracking data \cite{eye-tracking-system}.
In driving safety research, many crashes occur because drivers do not see the right object at the right time \cite{Eye-Tracking1}.
Eye tracking technology can provide real-time insights into a driver's attention, allowing the system to adapt the display of information based on where the driver is looking, avoiding cognitive overload. 
This could enhance the relevance and effectiveness of the information presented, ensuring that critical alerts are delivered in a way that is most likely to be noticed and acted upon by the driver.

Current research work shows that Driving Assistant Systems are good at identifying and warning risks on the road, but still face challenges in sensing and alerting risks from the roadside. 
How to design appropriate HUD interface for roadside risk warning needs further exploration.
\section{Visionary Co-Driver Enhancing Driver Perception of Potential Risks with HUD and LLMs}

\begin{figure*}[h!]
    \centering
    \includegraphics[width=1\linewidth]{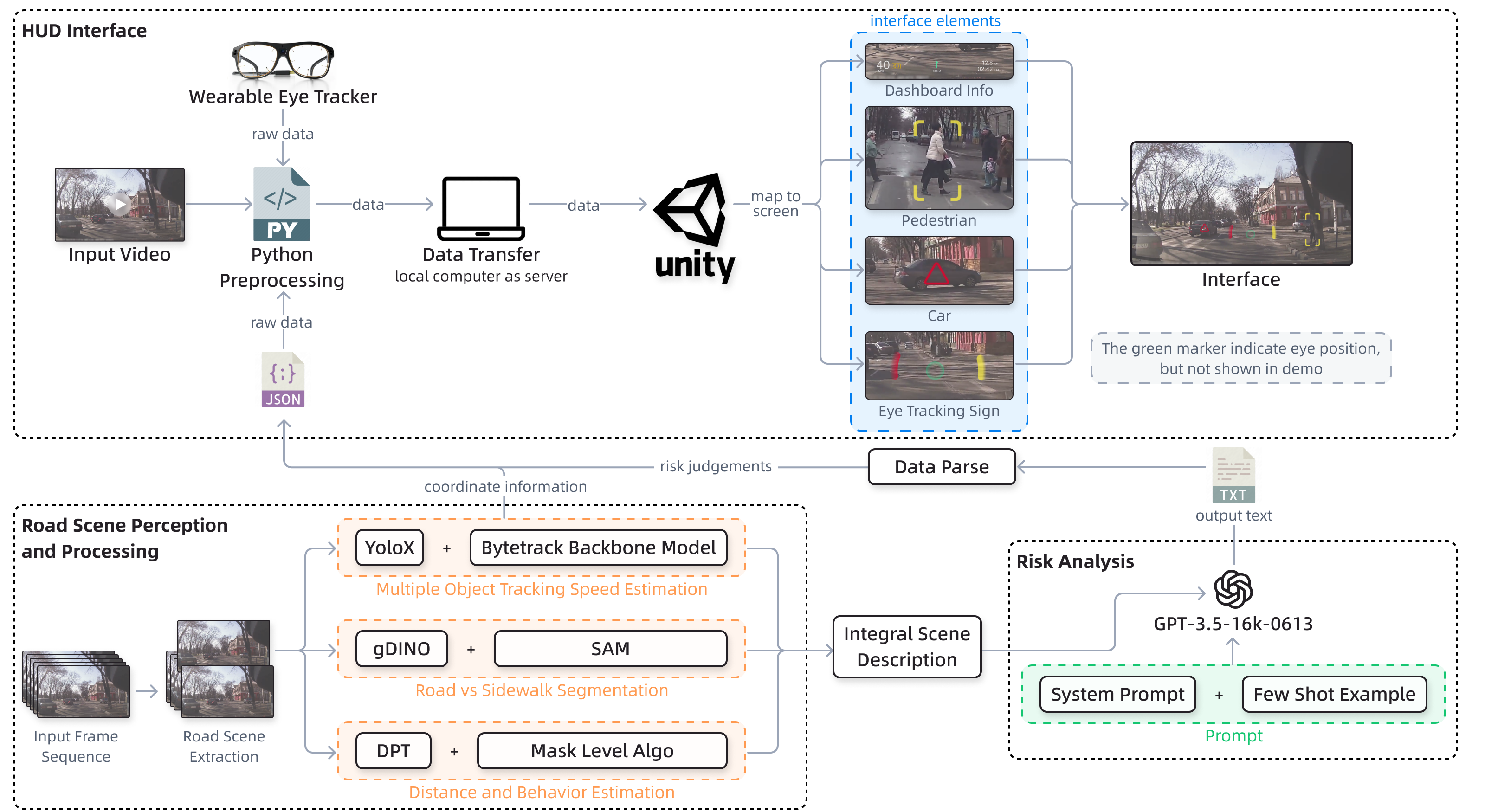}
    \caption{System design of VCD}
    \label{fig:system pipeline}
\end{figure*}

The ultimate goal of VCD is to enhance drivers' risk perception. 
It processes real-world visual data with computer vision technologies, identify risks from roadside road users by employing LLMs as a co-driver, and generates a visual warning in HUD, ensuring a safer and easier automated driving experience.

The VCD mainly consists of 3 modules, illustrated in Figure \ref{fig:system pipeline}:
\begin{enumerate}
    \item Road Scene Perception and Processing: Extract road scene information, such as position, depth and speed into natural language descriptions suitable for LLMs.  
    \item Risk Analysis: Utilize LLMs to analyze pedestrian intentions and relative directions to predict potential actions, and perform risk assessments. 
    \item HUD Interface: Translates LLMs' textual outputs into non-distracting visual warnings, effectively informing drivers of roadside risks.
\end{enumerate}

\subsection{Road Scene Perception and Processing}

\begin{figure}[!h]
    \centering
    \includegraphics[width=1\linewidth]{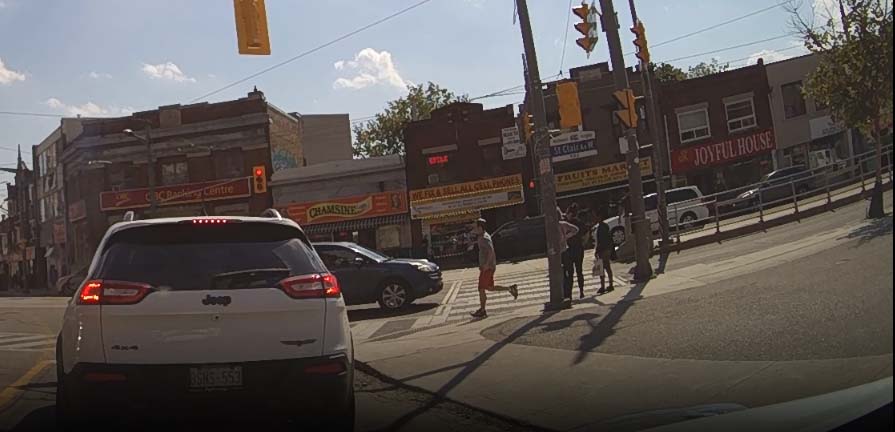}
    \caption{Typical potential roadside risk caused by pedestrian}
    \label{fig:scene_2}
\end{figure}

As shown in Figure \ref{fig:scene_2}, the input videos come from dash cam installed in vehicles.
In the example scene, crossroads and various urban streets represent typical environments of potential risks, with pedestrians possibly ignoring traffic regulations to cross the roads.

To obtain scene information on the road, we designed the following hybrid frame analysis framework.

\begin{enumerate}
    \item To enable LLMs with consistent and continuous reasoning on the image sequence, a Multiple Object Tracking (MOT) backbone model is indispensable. After evaluating various models, we integrated the YOLOX(You Only Look Once X) with ByteTrack \cite{yolov8, bytetrack} to associate low-confidence detection results for better tracking in crowded pedestrian environments.

    By integrating YOLOX with ByteTrack, our system extracts bounding boxes (bboxes) and assigns globally unique IDs to each pedestrian and vehicle in the video stream. 
    The extracted data, including bbox coordinates and unique IDs, are then formatted into JSON, ensuring seamless integration with the HUD interface.
    
    By analyzing the changes in the same pedestrian's bbox coordinates between different frames, we can anticipate their relative speed to the ego car, categorizing them into [``slow'', ``fast''].

    \item In complex urban traffic scenarios, determining whether pedestrians are stationed on the roadside is pivotal for intention estimation and risk assessment.

    To acquire more accurate data on the relationship between pedestrians and the road position, we utilized the pixel-level segmentation of Segment Anything Model (SAM) and Grounding DINO \cite{gdino} to distinguish sidewalks from roadways \cite{sam}, as Figure \ref{fig:all}(a) and Figure \ref{fig:all}(b) shows.

\begin{figure}[htbp]
    \centering
    \subfloat[]{\includegraphics[width=0.22\textwidth]{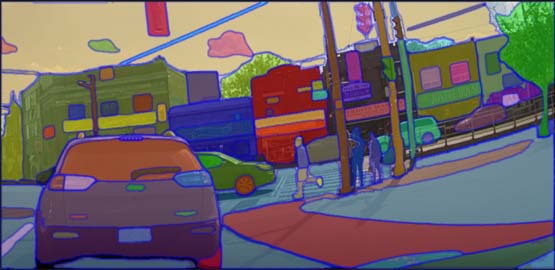}\label{fig:a}}
    \hspace{1pt} 
    \subfloat[]{\includegraphics[width=0.22\textwidth]{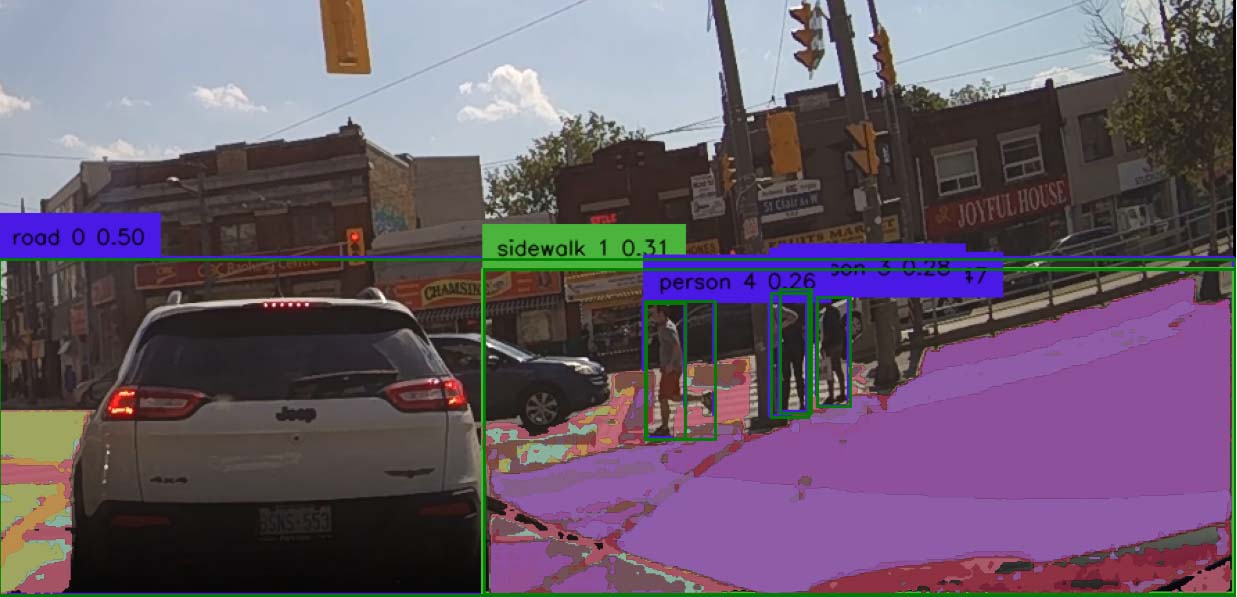}\label{fig:b}} \vspace{-3pt}\\
    \subfloat[]{\includegraphics[width=0.22\textwidth]{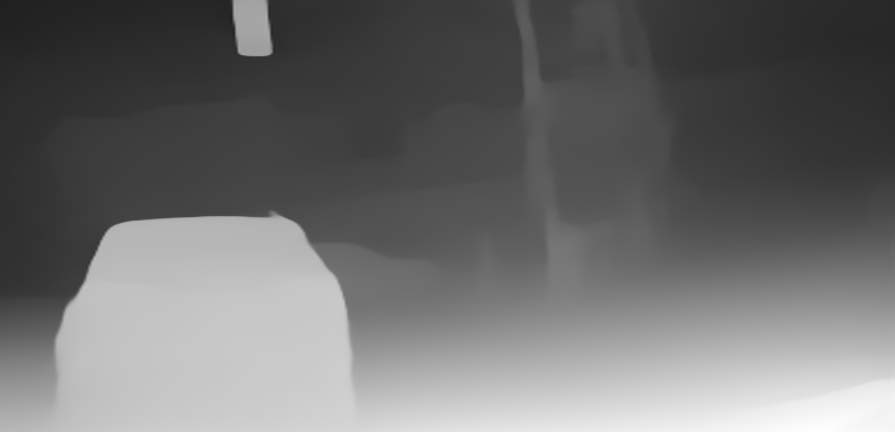}\label{fig:c}}
    \hspace{1pt} 
    \subfloat[]{\includegraphics[width=0.22\textwidth]{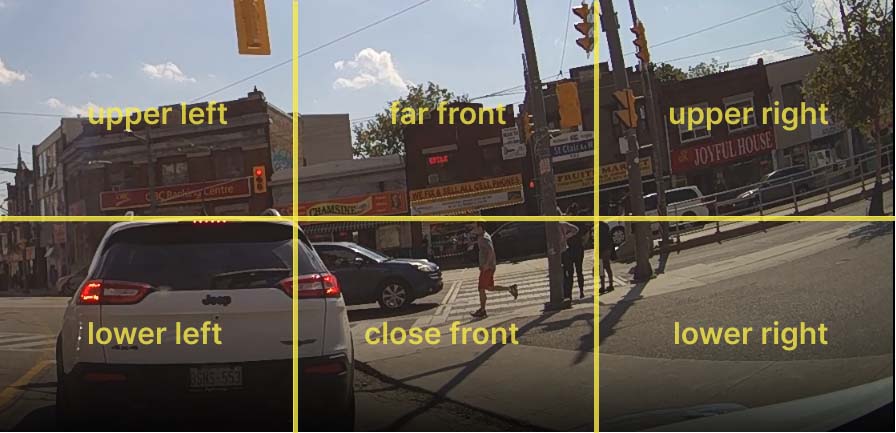}\label{fig:d}}
    \caption{Intermediate results of the frame analysis framework}
    \label{fig:all}
\end{figure}


    \item To obtain more road scene information from monocular video, we deployed the Dense Prediction Transformer (DPT) \cite{dpt} model for predicting the depth of each pixel in the current image, as Figure \ref{fig:all}(c) demonstrates. Pedestrian distance information, paired with the “safe distance” labels set based on traffic engineering principles, offers more contextual information and warnings to LLMs:
    \begin{enumerate}
        \item High Alert Area (0-5 meters): very near
        \item Alert Area (5-10 meters): near
        \item Caution Area (10-15 meters): medium
        \item Others: far and very far
    \end{enumerate}
    Given the distance and roadside masks generated by SAM, we apply a pixel-level algorithm to all of the pedestrians, obtaining their location and the type of surface they are currently on. As depicted in Figure \ref{fig:all}(d), we categorize sidewalks into ``left and right'' types, and roads into``left, center, right'' types. The pedestrian’s location is determined by dividing the picture into six sections, which are recorded as ``lower left, front center, lower right, upper left, far front, and upper right''. This process converts the pedestrian's bbox into location information that is easy for LLMs to understand, thereby reducing the complexity of the input information. 
    \item Data from various modules is aggregated and translated into an ``Integral scene description'' in natural language, which is then fed into the LLMs.
    
    Through this hybrid frame analysis, the LLMs can receive the following types of scenario:
    \begin{enumerate}
        \item The globally unique id of vehicles and pedestrians throughout the video.
        \item The bounding box trajectory.
        \item The detection confidence.
        \item The type of surface upon which the pedestrian is located.
        \item The location of pedestrian.
        \item The spatial orientation of vehicles and pedestrians, including distance and angle, relative to the ego vehicle.
    \end{enumerate}
\end{enumerate}

\subsection{Risk Analysis}
LLMs with extended context windows is favored since the risk reasoning \cite{chen2023extending, llm2023parallel}.
 is a complex task involving math and visual skill. VCD utilizes gpt-3.5-turbo-16k-0613 as logical kernel \cite{openai2023gpt35}, because it optimizes balance between the reasoning ability, memory length and economic efficiency.

The LLMs perform:

\begin{enumerate}
    \item Pedestrian intention analysis to predict actions such as potential street crossings and their relative direction to the ego vehicle.
    \item A comprehensive risk assessment, taking into account factors such as complexity and distance, to highlight areas of immediate concern within the current frame range.
    \item Binary classification of pedestrian behaviors to generate risk warnings for HUD.

\end{enumerate}

\begin{figure*}[h!]
    \centering
    \includegraphics[width=1\linewidth]{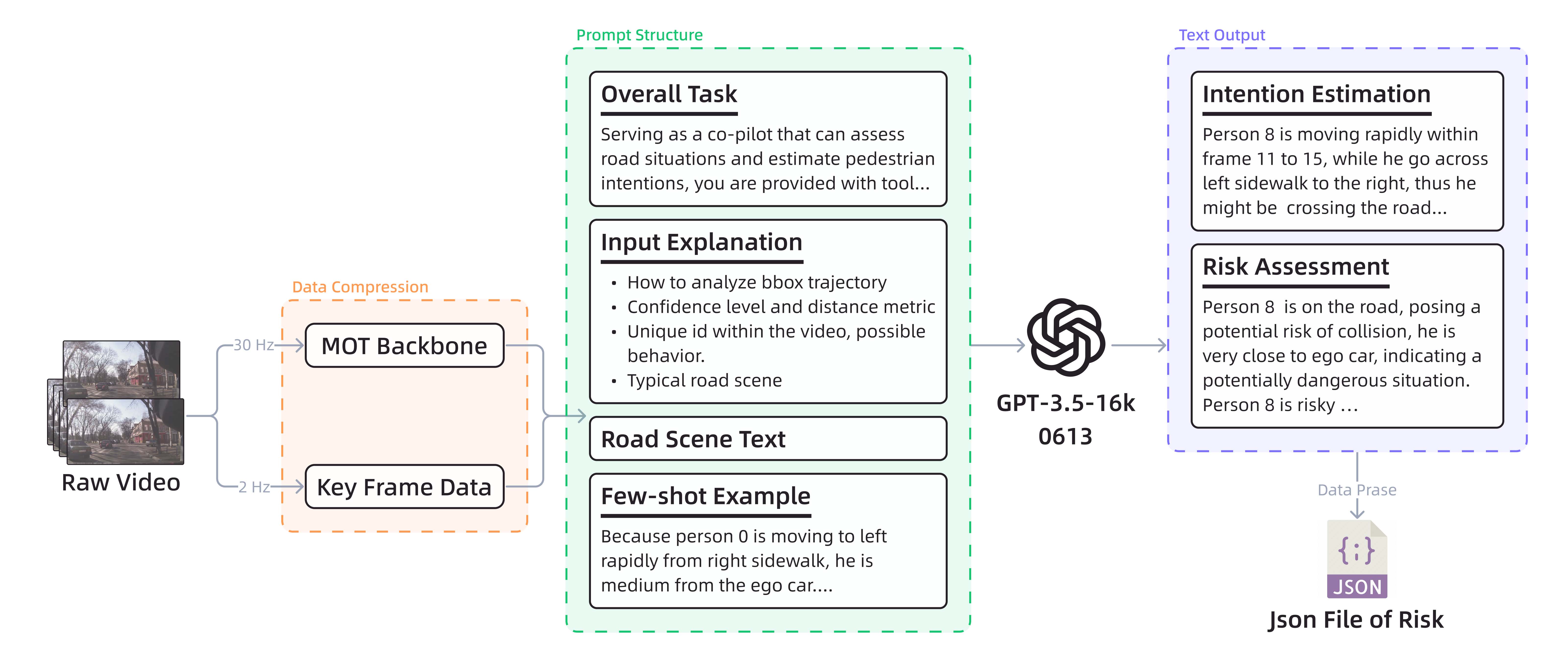}
    \caption{Reasoning process of VCD}
    \label{fig:gpt_think}
\end{figure*}

Figure \ref{fig:gpt_think} displays the reasoning workflow of VCD to realize risk analysis.

As shown in ``Prompt Structure'', the prompt template includes:
\begin{enumerate}
    \item Overall Task: inform the role of LLMs; 
    \item Input Explanation: description of the incoming data and some basic rules;
    \item Road Scene Text: road scene description converted from key frame image; 
    \item Few-shot Example: a few human-assisted reasoning examples.
\end{enumerate}
The prompt is then fed into GPT-3.5-16k-0613, facilitating intention estimation and risk assessment procedures. 
The final output portrays a structured JSON format encapsulating all recognized risky ids.

\subsection{HUD Interface}
\begin{figure}[h!]
    \centering
    \includegraphics[width=1\linewidth]{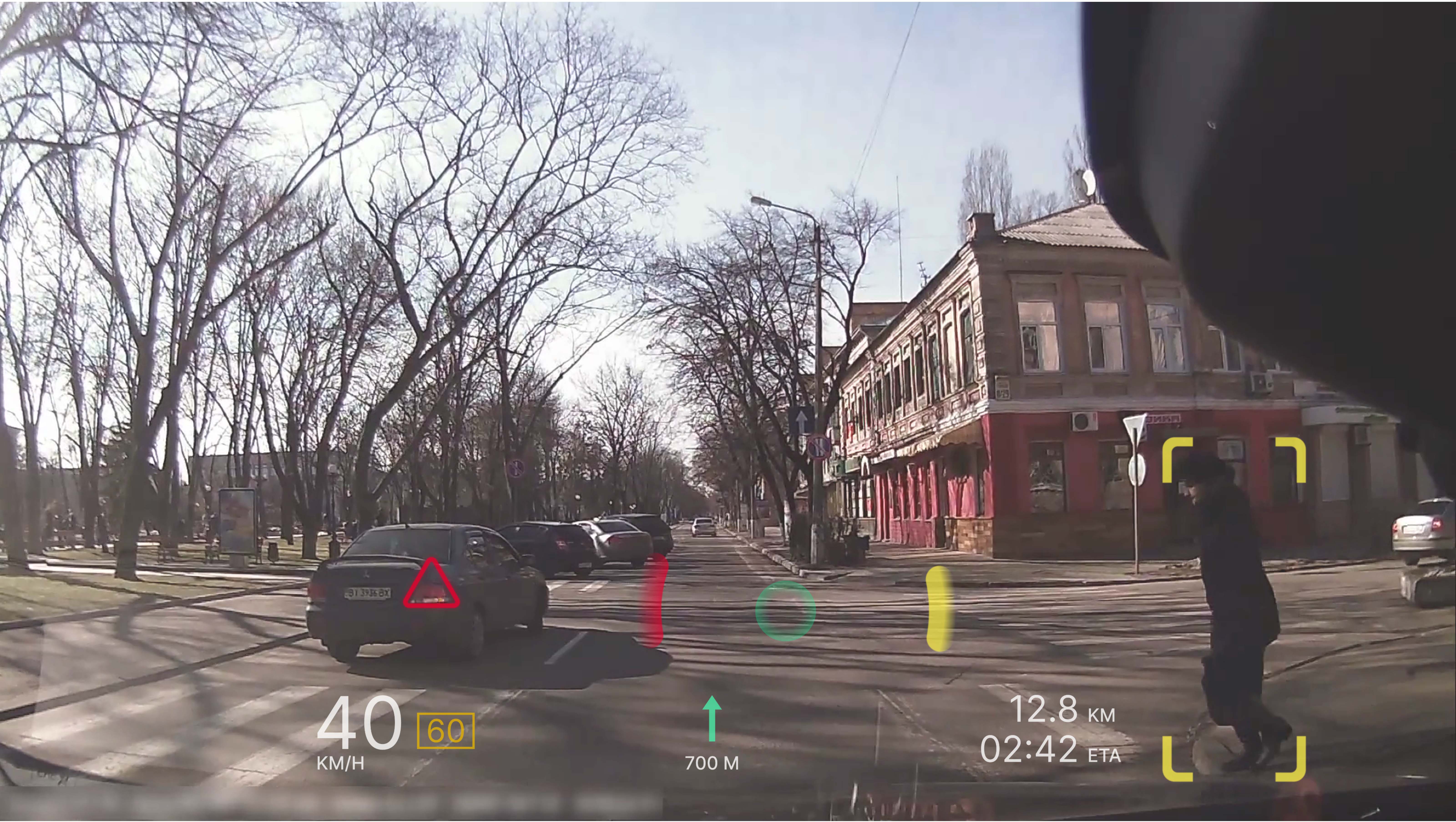}
    \caption{HUD interface}
    \label{fig:HUD Design Presentation}
\end{figure}

\begin{figure*}[h!]
    \centering
    \includegraphics[width=0.8\linewidth]{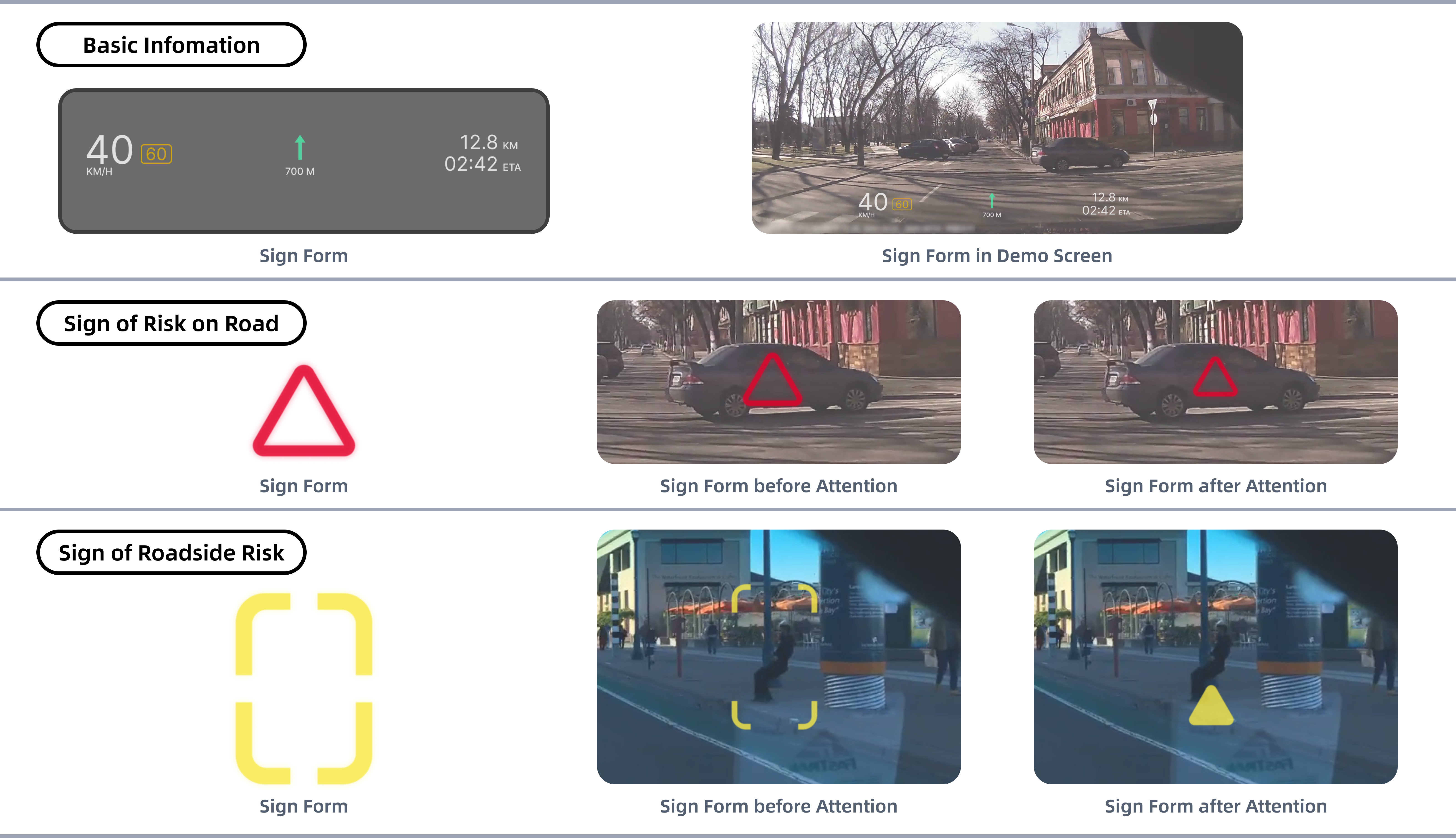}
    \caption{Risk sign form example}
    \label{Risk Sign}
\end{figure*}

Considering LLMs inherent uncertainties and risks of hallucinations, the interface is designed to provide visual cues with minimal distractions. 

The interface (Figure \ref{fig:HUD Design Presentation}) consists of three elements:
\begin{enumerate}
    \item Basic Driving Data: Basic Driving Data takes inspiration from WayRay \cite{wayray}, including speed, time and navigation direction.
It is positioned at the bottom of the interface without obstructing critical visual pathways, as shown in the top row of Figure \ref{Risk Sign}.
    \item Risk Signs: Risk signs including on-road risk signs and roadside risk signs. To effectively notify drivers about risks, we have tailored visual warnings for these two categories of risks as Figure \ref{Risk Sign} shows:
    \begin{enumerate}
        \item For risks on the road, a combination of red color and hollow triangles aims to create an obvious alert effect. The signs automatically shrink when the driver's sight is detected to be in the range of the sign by eye-tracking.
        \item For risks from roadside, bright yellow outline of a rectangle with corners only is used to prevent visual occlusion and maintain situational awareness. When the driver's sight is detected by eye tracking to enter the sign's range, the box outline disappears and turns into a small solid yellow triangle. If multiple adjacent targets (e.g., pedestrians) need to be labeled simultaneously, they will be merged into a single bounding box to reduce redundancy and to avoid excessive visual overload.
    \end{enumerate}
    \item Eye-guidance Signs: Two types of eye-guidance signs in the shape of arcs, pointing in the direction of the risk of road or roadside. Eye-guidance signs have the same color as the corresponding risk sign. These signs are designed as arcs, guiding the driver's gaze towards potential threats. We designed the eye-guidance signs to be positioned at the periphery of the participant's field of view, ensuring they serve as subtle directional cues without obstructing the driver's view or affecting their ability to perceive the driving environment.
\end{enumerate}

\begin{figure*}[h!]
    \centering
    \includegraphics[width=0.8\linewidth]{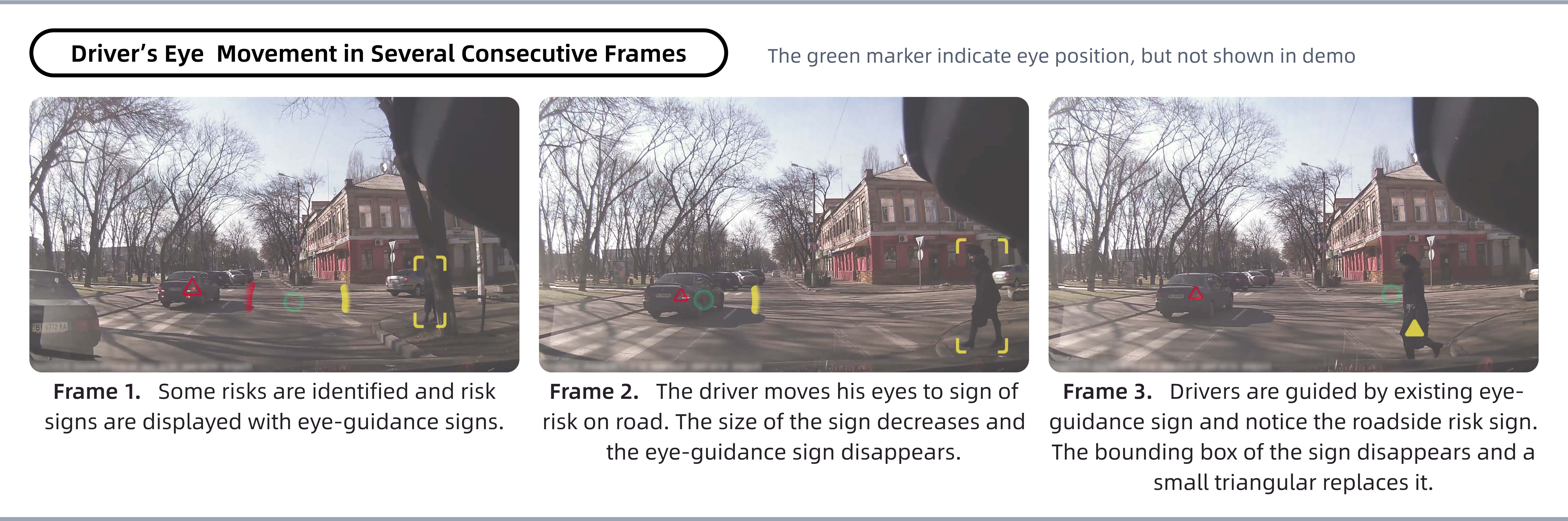}
    \caption{Eye-guidance sign}
    \label{fig:Eye-guidance sign presentation}
\end{figure*}

As shown in Figure \ref{fig:Eye-guidance sign presentation} ,
upon detection of potential risk, VCD poses visual signs , providing continuous guidance. 
After the driver focuses on a sign for potential risk, the corresponding on-road risk sign would turn smaller, the corresponding roadside risk sign would turn into small triangle, and the corresponding arc of eye-guidance sign would disappear to highlight risks that are unnoticed.
This dynamic adaptation is designed to enhance driver's risk perception without overwhelming their cognitive and visual faculties.

\section{Feasibility Evaluation Of VCD}
This section assesses the feasibility of the VCD system, focusing on two key aspects: correctness and response time. To evaluate correctness, we conducted metric evaluations on the output results from both the Road Scene Perception and Processing Module and the Risk Analysis Module. For response time, we performed detailed time measurements across several components: i) multiple models within the Road Scene Perception and Processing Module, ii) GPT-3.5 for intention estimation and risk assessment, and iii) the integration of eye-tracking with outcomes from the LLMs.
All tests were performed by default on an RTX 3090 graphics card.

\subsection{Correctness Evaluation}

\subsubsection{Road Scene Perception and Processing Module}
We evaluated the correctness of the Road Scene Perception and Processing Module in object detection, object tracking, and semantic segmentation of road and sidewalk. 
Given that the DPT model, utilized for depth estimation, has demonstrated strong performance on the ADE20K and Pascal Context datasets, even with smaller datasets yielding excellent results, we did not conduct additional evaluations \cite{dpt}.

\begin{table}[h]
\centering
\caption{Performance Metrics for the YOLOX-x Model}
\label{tab:yolo}
\begin{tabular}{c c c c c c c }
\toprule
\textbf{mAP50} & \textbf{TP(\%)} & \textbf{FP(\%)} & \textbf{FN(\%)} & \textbf{TN(\%)} & \textbf{Prec(\%)} & \textbf{Rec(\%)}\\ 
\midrule
55.6           & 51.9            & 9.8             & 38.3            & 0.0             & 84.1             & 57.5            \\ 
\bottomrule
\end{tabular}
\end{table}

    Object Detection. We assessed the models' metrics using subsets from the Joint Attention in Autonomous Driving (JAAD) dataset, comprising approximately 3,000 annotated sequences \cite{jaad}. 
    
    The YOLOX-X model was assessed using standard object detection metrics, including mean Average Precision at an Intersection over Union (IoU) threshold of 0.5 (mAP50), precision, recall, true positive (TP) rate, false positive (FP) rate, true negative (TN) rate, and false negative (FN) rate. Among these, mAP50 serves as a comprehensive metric that reflects both detection accuracy and robustness. Precision quantifies the proportion of correct detections among all predicted positives, while recall indicates the percentage of actual targets correctly identified by the model.

    The inference process for the YOLOX-X variant was conducted under the following parameter settings:
    \begin{itemize}
        \item A confidence threshold of 0.25 was applied to filter out low-confidence predictions, ensuring only reliable detections were retained.
        \item A non-maximum suppression (NMS) threshold of 0.45 was used to suppress overlapping bounding boxes, thereby reducing redundant detections.
        \item Input images were resized to a fixed resolution of 640×640 pixels to ensure standardized input dimensions, facilitating consistent processing and faster inference.
    \end{itemize}

    The performance of the YOLOX model on the JAAD dataset is summarized in Table (\ref{tab:yolo}).
    The results show that YOLOX-X achieves a relatively high precision of 84.1\%. Additionally, the model exhibits a low FP rate of 9.8\%, further demonstrating its reliability in avoiding incorrect detections.

    However, the FN rate remains relatively high at 38.3\%. This can be attributed to two primary factors: first, the JAAD dataset contains numerous distant pedestrians with fine-grained annotations, which pose a challenge for the original YOLOX architecture, these distant pedestrians are often relatively safe and not within the critical focus of our task, which means their impact on overall performance is minimal; second, the model was not specifically fine-tuned on diverse weather conditions present in the dataset, resulting in suboptimal generalization in adverse environments.
    It is also worth noting that the true negative rate is zero, which is expected in object detection tasks where background regions are not explicitly predicted.
    
    In summary, while there is room for improvement—particularly in reducing missed detections—the YOLOX-X model demonstrates a solid baseline performance on the challenging JAAD dataset. Its ability to run efficiently on commercial GPUs and achieve real-time inference speed makes it a promising solution for practical deployment. Fine-tuning the YOLOX-X model on the JAAD dataset is expected to significantly enhance its performance.
    
    Object Tracking. To evaluate the performance of the bytetrack\_x\_mot17 model, several key metrics were measured: Multiple Object Tracking Accuracy (MOTA), Identity F1 Score (IDF1), and Higher Order Tracking Accuracy (HOTA). MOTA evaluates the model's ability to accurately track multiple objects across frames, with higher scores indicating better overall tracking accuracy. The IDF1 score measures the effectiveness of maintaining consistent identities for tracked objects over time, reflecting both precision and recall in identity management. HOTA provides a comprehensive assessment of both detection and association precision, offering insight into the model’s capability to correctly identify and follow targets through complex scenarios.

    The bytetrack\_x\_mot17 model was configured with a customized set of parameters to optimize tracking performance. Specifically:
    \begin{itemize}
        \item A detection confidence threshold of 0.5 was applied to ensure that only high-confidence detections contributed to the tracking process.
        \item A track buffer of 30 frames was used to temporarily retain tracks that may have been occluded or missed, thereby accommodating brief interruptions in detection.
        \item A matching threshold of 0.8 was defined as the minimum overlap required between a detection and an existing track for them to be associated.
        \item An aspect ratio threshold of 1.6 was employed to filter out detections inconsistent with typical pedestrian proportions, reducing false positives. Additionally, a minimum bounding box area of 10 pixels was enforced to eliminate overly small detections that could represent noise or background artifacts.
        \item The ground truth data was preprocessed to include only ``nearby'' and ``distant'' pedestrian labels, excluding the ``crowd'' category, in order to focus on individual tracking performance.
    \end{itemize}

    Under these settings, the bytetrack\_x\_mot17 model achieved a MOTA of 74.7\%, an IDF1 score of 80.3\%, and a HOTA of 73.1\%. These results reflect strong tracking accuracy, robust identity preservation, and reliable association performance, demonstrating the model's effectiveness in handling complex, real-world multi-object tracking scenarios.

    Road and Sidewalk Segmentation. We extracted 100 driving scene images from the Berkeley DeepDrive (BDD) dataset and used the Grounding DINO + SAM algorithm module to segment the images, obtaining masks for persons, sidewalks, and roads within the images \cite{bdd}. We invited a human expert with six years of driving experience to assess and evaluate the accuracy of the segmentation. Specifically, the segmentation accuracy for the person category was 78.0\%, for the sidewalk category it was 54.8\%, and for the road category, it reached 97.0\%, indicating an relatively high segmentation accuracy.

\subsubsection{Risk Analysis Module}
Existing datasets primarily offer explicit, objective indicators (such as whether the pedestrian crossed the street, the pedestrian motion state, and the first frame of the driver's response \cite{RoadHazardStimuli,bdd}). 
However, existing datasets typically do not provide risk assessment in non-collision situations. 

Taking these factors into account, we define a ``reasonable'' judgment as one that strikes a balance between ``aggressive'' and ``conservative'' driving.
If the system does not point out what the human expert thinks should be warned, it is considered ``aggressive''. If the system points out what the human expert thinks is unnecessary, it is considered ``conservative''.
Our evaluation strategy involves collecting expert assessments of the VCD's judgments across the dimensions of ``aggressive / conservative''. 
The degree of aggression or conservatism within a judgment reflects the level of its reasonableness.

We invited two experts (average driving experience 5.5 years) to verify the credibility of VCD judgments.

We randomly select 50 videos from JAAD dataset as validation sets\cite{jaad}.
JAAD dataset provides a richly annotated collection of 346 short video clips (5-10 seconds long) extracted from over 240 hours of driving footage\cite{jaad_annotation}. These videos represent scenes typical of everyday urban driving in various weather conditions.
The VCD makes judgments about whether a pedestrian in the current second is safe or risky based on the information from the previous three seconds. 
Then pedestrians judged to be risky are highlighted in the videos according to the second-by-second assessment.
The experts were asked to watch the highlighted videos and rate the VCD's judgments on each video on a seven-point scale, with one being extremely conservative and seven being extremely aggressive.

After verifying the consistency of the two experts' ratings using a two-way random effects model (\textit{ICC} = .61, \textit{p} \(<\) .01), we obtained an average score of 3.07 for the two experts indicating that the judgments tend to be reasonable and slightly conservative.
\subsection{System Latency Analysis}
\label{sec:latency}
To analyze the run-time of each sub-module in the VCD, we selected 50 video samples from JAAD\cite{jaad} and tested each module's delay on RTX 3090.

To fairly benchmark each module, we clipped each video sample to the first 2 seconds, downsampled from 30Hz to 2Hz, and obtained 4 RGB frames as a casual window input for the VCD, calculating the average latency.

\begin{table*}[t]
    \centering
    \caption{Module-level latency for a 2 s causal window (2 Hz, 4 RGB frames)}
    \label{tab:module_latency}
    \begin{tabular}{@{}p{0.27\textwidth}p{0.65\textwidth}@{}}
        \toprule
        \textbf{Module} & \textbf{Measured latency (per clip)} \\
        \midrule
        1. Road Scene Perception and Processing & 1) Object tracking: 0.19s, with 0.07s of YOLOX-X and 0.12s of byte\_track\_x\_mot17 \\
         & 2) Scene segmentation and crounded Captioning: 3.86s, with 0.46s of gDINO and 3.40s of SAM1 \\
         & 3) Depth estimation: 14.97s of DPT \\
        2. Risk Analysis & 1.76s with GPT-3.5-16k \\
        3. HUD Interface & 27--33 ms, mainly from Tobii eye-tracking device \cite{TobiiLatencyZH} \\
        \bottomrule
    \end{tabular}
\end{table*}

As shown in Table \ref{tab:module_latency}, 3 listed pipeline of Road Scene Perception and Processing can be executed in parallel; therefore, the bottleneck is determined by the longest stage among them. 

In our initial completion of this work, we believed that the delay would rapidly decrease as technology advances. Using the earliest version of the project (Table \ref{tab:module_latency}), the latency was 16.76s for 2 seconds video clip. 
However, by using more advanced models and methods — for example, a more efficient SAM2 model with a latency of 0.10 s for grounded captioning, replacing the depth estimation module with LiDAR/Radar depth assessment (0.20 s), and employing the lightweight GPT-4.1-nano model (1.40 s) for risk analysis — we can reduce the total latency for processing a 2-second video clip to 2.29 seconds (calculated as: 0.56 s for Module 1 + 1.40 s for Module 2 + 0.033 s for Module 3) \cite{sun2024efficientvariantssegmentmodel,velodyne2019vls128, continental2025ars540}.

Current driver assistance systems typically use TTC thresholds for emergency collision in the range of 1.5–4 seconds \cite{MATTAS2020105794}. Therefore, the VCD can provide advice while the driver still has more than 1.5 seconds to react to a potential risk. With ongoing technological advancements, the latency is expected to decrease further, making the system increasingly suitable for real-time applications.

Performing live on-road experiments with an early-stage prototype is unsafe and poses challenges for replication and validation. Therefore, we utilize video recordings to facilitate the testing of both human subjects and the eye-tracking HUD in a controlled laboratory environment.
Another reason for choosing the video-based approach is that our focus was on understanding users’ recognition of risk — a key aspect that can be effectively captured through video-based observation. Although this method may not fully replicate the realism of real-world driving, it provides valuable insights while ensuring safety and experimental control. Additionally, participants could also simulate driving operations while watching the videos, which allowed us to gather meaningful data without compromising safety.

\section{User Study}

This study\footnote{University ethics review board approves human-subjects research and they approved this project (2024096).} examines the effectiveness of VCD by testing driver's risk perception, workload, and user experience. We have the following hypotheses:

\begin{itemize}
    \item H1: VCD can improve driver's risk perception of roadside road users.
    \item H2: VCD does not compromise the driver's risk perception on the road.
    \item H3: Compared with the HUD without roadside risk warning, the use of VCD will reduce the cognitive load of the driver.
\end{itemize}

\subsection{Apparatus}
\begin{figure}[!h]
    \centering
    \includegraphics[width=1\linewidth]{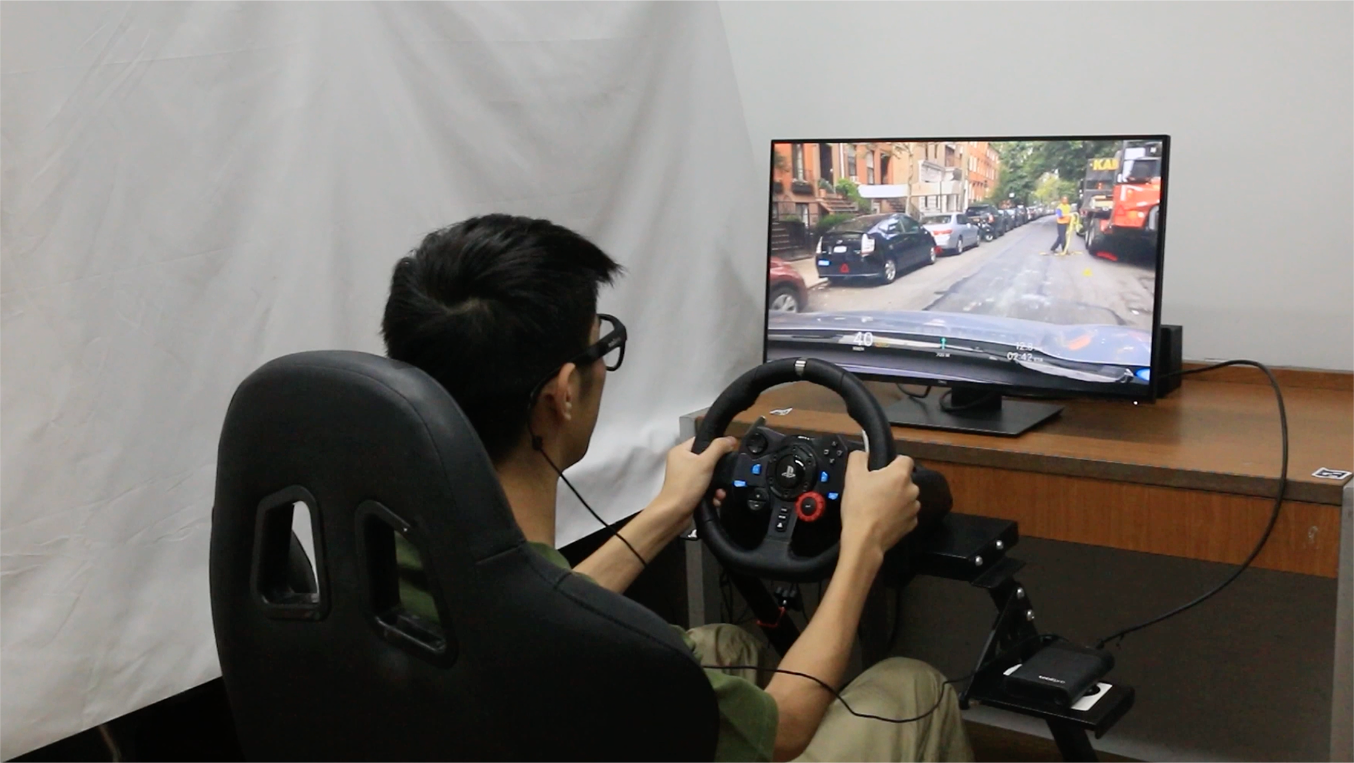}
    \caption{The driving simulator setup}
    \label{fig:test scene}
\end{figure}

The driving simulator setup (Figure \ref{fig:test scene}) included a cockpit, a DELL U2720Q 27-inch LED screen, and a Tobii Pro Glasses 3 portable eye-tracking device. The vertical distance between the LED screen and the participants was about 0.8 meters. All virtual scenes in the experimental setup were processed with Unity and displayed on LED screens. Tobii Pro Glasses 3 recorded participants' eye movement at a sampling frequency of 100 Hz. The eye-tracking data was then analyzed using Tobii Pro Lab.

\subsection{Experimental Metrics}
\begin{enumerate}
    \item Risk perception was measured by Situation Awareness Global Assessment Technique (SAGAT)\cite{sagat}. SAGAT effectively gauges risk perception within dynamic contexts, assessing individuals' awareness of potential threats and system statuses.
    \item Workload was measured in different conditions by Subjective Workload Assessment Technique (SWAT), which is a method used to measure the cognitive workload of human operators in real time\cite{swat}. It is a multidimensional scale, whose dimensions are time load, mental effort load, and psychological stress load.
    \item Participants' feelings about HUDs were measured by User Experience Questionnaire (UEQ) \cite{UEQ}. The UEQ is a comprehensive tool for measuring both usability and experiential aspects of interactive product use.
    \item Participants' views on HUD were obtained by semi-structured interviews.
\end{enumerate}

\subsection{Study Design}
\begin{figure*}
    \centering
    \includegraphics[width=1\linewidth]{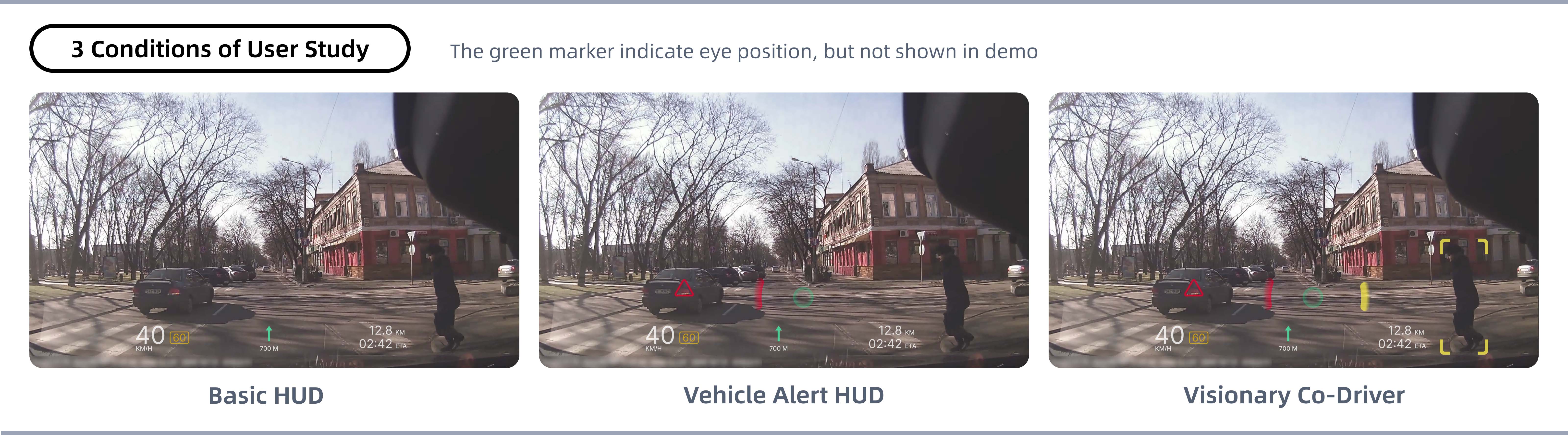}
    \caption{Three HUD conditions}
    \label{fig: Three conditions of HUD}
\end{figure*}

To test the effects of VCD, the study included three HUD conditions.
Figure \ref{fig: Three conditions of HUD} illustrates three HUD conditions.
\begin{enumerate}
    \item \textbf{Basic HUD condition}: In this condition, the HUD presents basic driving information including speed, time and navigation direction.
    \item  \textbf{Vehicle Alert HUD condition}: In this condition, the HUD presents red triangle signs to indicate vehicles that posed a threat to the ego vehicle.
    \item  \textbf{VCD condition}: In this condition, the HUD presents VCD interface.
\end{enumerate}

Our driving scenarios were regular daytime driving scenarios collected from BDD dataset and JAAD dataset.

In each condition, participants saw five video clips.
The order of HUD condition and the order of video clips were counterbalanced among all participants to reduce order and learning effect.

\subsection{Procedure}
\begin{figure*}
    \centering
    \includegraphics[width=1\linewidth]{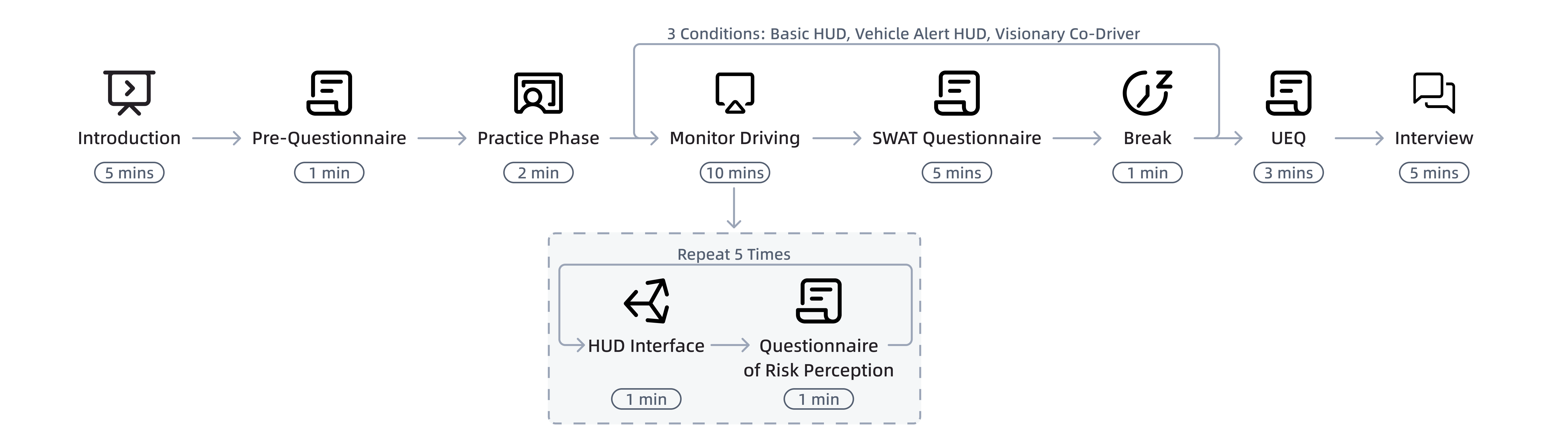}
    \caption{User study procedure}
    \label{fig:Procedure}
\end{figure*}

Figure \ref{fig:Procedure} illustrates the study's process flow. 
\subsubsection{Introduction Phase}
 Upon arrival, each participant signed a consent form confirming their voluntary participation. They were introduced to the experiment procedure (e.g., that they could quit the experiment at any time in case of motion sickness or any discomfort) and were told what they would experience during the study.

\subsubsection{Pre-Questionnaire Phase}
Participants were assisted in fitting and calibration of Tobii Pro Glasses 3.
Then, they completed a pre-questionnaire gathering their basic information.
Their eye movement was monitored through Tobii Pro Glasses 3 to ensure accurate positioning.
If participants moved or removed the glasses during the experiment, the glasses needed to be recalibrated.

\subsubsection{Practice Phase}
Participants were instructed to immerse themselves in the experiment from a first-person perspective.
They were asked to repeat the numbers given by the experimenter at a steady pace while monitoring the driving process.
The practice phase consisted of three practice trials, totaling approximately 2 minutes, featuring simplified road settings to help participants acclimate to the experimental environment and different HUD conditions.

\subsubsection{Formal Study Phase}
Participants entered the formal study phase, in which they would experience all HUD conditions (Basic HUD, Vehicle Alert HUD, VCD). 
Every condition included 5 video clips, whose difficulty was balanced across all HUD conditions by scoring and reordering video clips based on frozen time and scene complexity.
The video clips went blank at the selected frozen points so that participants filled out a risk perception questionnaire derived from SAGAT.
Following the viewing of video clips under a specific HUD condition, participants filled out an SWAT questionnaire.
After resting, participants switched to the next HUD condition.

\subsubsection{UEQ and Interview Phase}
Upon completing all experimental trials, participants were asked to complete the UEQ. The experimental conductor conducted semi-structured interviews about the study.

\subsection{Participants}
We recruited 41 licensed drivers with normal to corrected-to-normal vision. Due to the malfunction of Tobii Pro Glasses 3, we obtained valid data from 36 participants.
There were 17 (47\%) males and 19 (53\%) females.
With a mean age of 21.61 years (\textit{SD} = 2.49), participants averaged 1.72 years of driving experience (\textit{SD} = 1.53). Their mean annual mileage was 403.31 km (\textit{SD} = 449.64).
\section{Results}

\subsection{Risk Recognition of the Three HUD Conditions}
\begin{table*}
    \caption{Driver's accuracy of risk recognition in three HUD conditions}
    \centering
    \begin{tabular}{ccccccccc}
        \toprule
        HUD conditions      & \multicolumn{2}{c}{All Objects} & \multicolumn{2}{c}{Pedestrians} & \multicolumn{2}{c}{Vehicles} & \multicolumn{2}{c}{Environment Elements}                                                       \\
                            & \textit{M}                      & \textit{SD}                                 & \textit{M}                  & \textit{SD}               & \textit{M} & \textit{SD} & \textit{M} & \textit{SD} \\
        \midrule
        Basic HUD           & .60                             & .08                                         & .55                         & .14& .57        & .57         & .62        & .62         \\
        Vehicle Alert HUD   & .61                             & .09                                         & .60                         & .18& .58        & .59         & .62        & .62         \\
        VCD & .61                             & .07                                         & .67                         & .15& .57        & .57         & .60        & .60         \\
        \bottomrule
    \end{tabular}
    \label{tab:sagat_result}
\end{table*}

We scored the risk recognition questionnaire and calculated the accuracy of risk recognition (Table \ref{tab:sagat_result}).
The average accuracy for pedestrians was lowest in the Basic HUD condition (\textit{M} = .55; \textit{SD} = .14) , intermediate in the Vehicle Alert HUD condition (\textit{M} = .60; \textit{SD} = .18) , and highest in the VCD condition (\textit{M} = .67; \textit{SD} = .15).

A repeated measures ANOVA with the Bonferroni correction for multiple comparisons was used to determine whether there was an effect of different HUD conditions on risk perception.

HUD conditions did not have a statistically significant effect on the accuracy of risk recognition for All Objects (\textit{F} (2, 70) = .36, \textit{p} = .70, \textit{$\eta$\textsuperscript{2}} = .02).

HUD conditions did not have a statistically significant effect on the accuracy of risk recognition for Vehicle (\textit{F}(2, 70) = .17, \textit{p} = .85, \textit{$\eta$\textsuperscript{2}} \(<\) .01).

HUD conditions had a statistically significant effect on the accuracy of risk recognition for Pedestrian (\textit{F}(2, 70) = 4.83, \textit{p} = .01, \textit{$\eta$\textsuperscript{2}} = .12).

HUD conditions did not have a statistically significant effect on the accuracy of risk recognition for Environment Elements (\textit{F}(2, 70) = .46, \textit{p} = .64, \textit{$\eta$\textsuperscript{2}} = .01).

\subsection{Eye Tracking Data}
HUD conditions had little impact on eye-tracking data.
Repeated measures ANOVA was used to test eye movements across conditions.
We found no significant main effects for average duration of whole fixation (\textit{F} (2, 61) = 1.73, \textit{p} = .19, \textit{$\eta$\textsuperscript{2}} = .47), average whole fixation pupil diameter (\textit{F} (2, 70) = .94, \textit{p} = .40, \textit{$\eta$\textsuperscript{2}} = .03) or average amplitude of saccades (\textit{F} (2, 70) = .25, \textit{p} = .78, \textit{$\eta$\textsuperscript{2}} = .01).

\subsection{User Experience}
\begin{table}
    \caption{UEQ questionnaire results for both HUD conditions}
    \label{tab:ueq}
    \centering
    \begin{tabular}{ccccc}
        \toprule
                       & \multicolumn{2}{c}{Vehicle Alert HUD} & \multicolumn{2}{c}{VCD}                            \\
        Scale          & \textit{M}                            & \textit{SD}                             & \textit{M} & \textit{SD} \\
        \midrule
        Attractiveness & .56                                   & .81                                     & .70        & 1.22        \\
        Perspicuity    & 1.18                                  & 1.21                                    & .67        & 1.34        \\
        Efficiency     & .50                                   & 1.15                                    & .55        & 1.33        \\

        Dependability  & .54                                   & 1.02                                    & .51        & 1.31        \\
        Stimulation    & .66                                   & .94                                     & 1.13       & .89         \\
        Novelty        & .73                                   & 1.20                                    & 1.59       & 1.10        \\
        \bottomrule
    \end{tabular}
\end{table}

The results of the UEQ questionnaire are shown in Table \ref{tab:ueq}.

Users scored higher on Stimulation (\textit{F} (1, 35) = 5.02, \textit{p} =.03,  \textit{$\eta$\textsuperscript{2}} = .13) and Novelty (\textit{F} (1, 35) = 9.74, \textit{p} \(<\) .01,  \textit{$\eta$\textsuperscript{2}} = .22) with Visionary Co-Driving compared with the Vehicle Alert HUD.
Vehicle Alert HUD scored 1.18 for Perspicuity, .66 for Stimulation and .73 for Novelty.
Visionary Co-Driving scored .70 for Attractiveness, .67 for Perspicuity, 1.13 for Stimulation and 1.59 for Novelty.

\subsection{Cognitive Workload}

\begin{table}
    \caption{Cognitive workload in three HUD conditions}
    \label{tab:swat}
    \centering
    \begin{tabular}{ccc}
        \toprule
        HUD conditions      & \textit{M} & \textit{SD} \\
        \midrule
        Basic HUD           & 10.19      & 7.27        \\
        Vehicle Alert HUD   & 6.14       & 5.53        \\
        VCD & 9.86       & 7.06        \\
        \bottomrule
    \end{tabular}
\end{table}
As Table \ref{tab:swat} shows, HUD condition had no significant impact on cognitive workload.
The results of SWAT showed participants using Basic HUD scored 10.19, participants using Vehicle Alert HUD scored 6.14, and participants using VCD scored 9.86 (\textit{F} (2, 70) = 4.15, \textit{p} = .02, \textit{$\eta$\textsuperscript{2}} = .11).

\section{Discussion}
Comparing VCD with Basic HUD and Vehicle Alert HUD, our experiments verify the effectiveness in improving driver risk perception, providing novel user experience without excessive cognitive workload. This validates the benefits of LLMs engagement for drivers and offers the possibility of close collaboration between the two.

\subsection{The Effect of VCD on Risk Perception}
H1 is substantiated.
Compared with Basic HUD, VCD improved risk perception for pedestrians.
The risk recognition for pedestrians increases from Basic HUD, Vehicle Alert HUD to VCD.

H2 is substantiated.
There was no significant difference in risk perception for vehicles and other environment elements, so VCD does not affect participants' ability to recognize objects other than pedestrians.

It's proved that with LLMs' assistance, the driver assistance system excels at identifying roadside risks in non-collision situations.
While acknowledging the uncertainty and hallucinations in LLMs, we find that the judgments made by LLMs are reasonable and slightly conservative.
Therefore, it is preferable to employ the LLMs as an adjunct to driver's decision-making, ensuring that ultimate control remains firmly in the driver's hands, as our interface is designed to do. 
Considering the varying driving styles among drivers, the system can be fine-tuned to be more aggressive or conservative through prompt engineering or fine-tuning to cater to individual preferences.

\subsection{The Effect of VCD on Cognitive Workload}
H3 does not hold.
The subjective workload in the VCD condition was not significantly different from the Basic HUD and Vehicle Alert HUD conditions.
This may be due to the large individual variation among drivers, as indicated by the standard deviation.
However, we observed that the average workload score of Vehicle Alert HUD is lower than that of Basic HUD and VCD, which indicates that the icons added by Vehicle Alert HUD affect the driver's state.

\subsection{User Experience and Feedback}
VCD design performed better in Stimulation and Novelty.
The higher score in Stimulation suggests that our design effectively captures users' attention and piquing their interest.
Furthermore, a higher score in novelty implies that our design deviates from traditional or conventional solutions to some extent.
This differentiation could be attributed to unique design elements, innovative interaction methods, or unprecedented functionalities.
In conclusion, Visonary Co-Driver's performance in terms of Stimulation and Novelty positively influences user experience.
Users are more likely to be attracted to our design and enjoy interacting with it.
These findings support our design choices and decisions.

\subsection{The Cooperation between Drivers and LLMs}
VCD considers both the out-of-vehicle driving environment status and the in-vehicle driver status for risk warning.
It utilizes extensive prior knowledge from LLMs to analyze potential risks on the roadside and incorporates the user's eye movement data, displaying different risk signs with HUD based on the driver's focus. 

    \subsubsection{Rational use of LLMs in Human-Vehicle Cooperation:}
    In human-automation cooperation, it is crucial to consider the user's expertise and the system's time sensitivity and safety requirements \cite{HistoryandFuture}.

    Automated driving is used in a time-sensitive, safety-critical context. Drivers may also have a tendency to underestimate potential risks or place unwarranted trust in the automation, which can result in risky situations.
    This misalignment in expectations can cause misuse or disuse of the technology, where users either rely too heavily on it or avoid using it due to a lack of understanding.
    Given the unresolved challenges, we should design with a tolerance for error in mind. Accordingly, in VCD, we have implemented several designs:
    \begin{itemize}
        \item The system acts as an assistant rather than a decision maker for drivers, so that it provides reminders of risks and leaves the final decision remaining in drivers' hands.
        \item Considering the possible fault of the system, the system employs an intuitive and non-distracting interface which dynamically adjusts its visual presence in response to drivers' awareness of warning signs. In this way, the interface acts as a hint rather than a strong alert.
    \end{itemize}
    
    \subsubsection{Potential Applications as Co-driver Agents: }
    The reasoning ability of LLMs comes from large-scale pretraining, which has got more attention to construct a decision-making agent in diverse driving situations. As co-driver agents, such integration of LLMs and driver's state can play a multifaceted role in supporting drivers. Beyond pedestrian intention prediction, they can act as voice-activated passenger service providers, offering navigation assistance, answering queries, and tending to passengers' needs, enhancing the overall driving experience. This versatile approach of incorporating LLMs into co-driver systems and analyzing the driver's state holds potential for various applications, from advanced driver assistance to improved user interaction and risk perception.


\subsection{Limitations and Future Work}

While current developments in the integration of LLMs as a co-driver agent and HUD have shown great promise in enhancing the driving experience, several areas deserve attention for future work. This section highlights these limitations and suggests directions for future research and improvement in the field.

\subsubsection{Scenarios Where Scene Perception Alone May Be Sufficient}

       In certain driving scenarios, especially those with clear visibility and minimal distractions, the number of system prompts naturally decreases because there are fewer risks that need to be highlighted. This reduction not only minimizes distractions for the driver but also alleviates their cognitive load. Moreover, it is even possible that in these simple scenarios, drivers can easily detect and respond to potential risks on their own, rendering the additional complexity and computational overhead introduced by LLMs unnecessary.

However, the strength of LLMs lies in handling complex and uncertain scenarios where scene perception alone may fall short. 
For example, when multiple pedestrians with varying intentions and movements are present on the roadside, the driver's cognitive load can increase significantly. 
In such complex scenarios, LLMs can provide valuable assistance by analyzing the collective behavior of pedestrians and identifying potential risks that the driver might overlook due to information overload.

It is worth noting that there exists a threshold beyond which additional prompts may no longer be beneficial and could even increase the driver's burden. Future work will focus on exploring this threshold to achieve better adaptive adjustments, ensuring that system prompts enhance safety without becoming burdensome.

\subsubsection{Comprehensive Real-world Testing}

          The experimental scenarios used to evaluate LLM-based co-driver agents and HUD are limited in their scope.
          To ensure the robustness and reliability of these systems, it is imperative to include real driving scenarios, including corner cases and challenging driving situations.
          Future work should involve comprehensive test scenarios and more diverse participants that challenge the capabilities of LLMs-based HUD system in various real-world situations.
          This can include adverse weather conditions, sudden road obstacles, and complex traffic scenarios.

\subsubsection{Multimodal Alerts}

          There are numerous visual blind spots in driving, and these areas may have suboptimal effectiveness for visual warning, which can be supplemented by auditory cues.
          In addition, means such as touch and smell can also be considered.
          By utilizing multisensory channels, more comprehensive and accurate information can be provided, aiding users in understanding and adapting to specific situations.
          Therefore, multimodal warnings serve as a design point for enhancing the interface in future iterations.

\subsubsection{Participants Selection}
The main participants in this study are young drivers, for whom the effects of our system may be pronounced. Specifically, according to data from the National Highway Traffic Safety Administration (NHTSA), drivers under the age of 25 have an accident rate 1.6 times higher than that of other age groups\cite{lombardi2017age}. Additionally, young drivers tend to exhibit stronger adaptability to new technologies and display higher levels of behavioral plasticity, enabling them to adjust and improve more quickly when using the system.

Nevertheless, we recognize the limitation of our current participant pool. To improve generalization of our findings, future studies could extend the evaluation to a more diverse range of drivers, spanning different age groups (e.g., middle-aged and older adults), occupational backgrounds (e.g., professional drivers), and levels of driving exposure (e.g., low- vs. high-mileage drivers). This would enable a more comprehensive assessment of the system’s performance in varied user profiles and real-world usage patterns.

\subsubsection{LLM-based Risk Analysis}

Despite their great potential for risk detection, LLMs may not match the processing speed of the fastest existing methods. Its inference time is proportional to the scene complexity, while this could benefit with the improvement from quantization, tensor rt transformation, parallelization, etc. Meanwhile, it's challenging for one universal LLMs to generalize across cultural contexts, as differences in traffic rules between countries could introduce more complexity for reasoning. 
However, LLMs possess two irreplaceable advantages. First, they can integrate multimodal cues by combining visual, auditory, and other types of information to achieve a more comprehensive understanding of traffic scenarios. Second, LLMs exhibit strong scenario adaptability through their ability to process and understand natural language instructions and context. This enables LLMs to dynamically adapt their responses based on new feedback and evolving traffic conditions, offering a level of flexibility and context-awareness that traditional methods struggle to match. This adaptability is particularly valuable in complex and dynamic traffic environments where rules and conditions can vary significantly.

\subsubsection{Towards Safe and Responsible Deployment}
While the LLM serves only as a cognitive assistant, its potential for hallucination remains a concern for real-world deployment.

To enhance output reliability, real-world deployment should incorporate mechanisms to validate and bound the LLM's outputs.
Such mechanisms could include plausibility checks based on object motion and distance estimates, consistency verification across consecutive frames, and suppression of alerts when the LLM's output confidence is low.
For example, if the LLM identifies a distant pedestrian on the sidewalk as high risk but tracking data shows no movement toward the road, the alert can be downgraded.
Similarly, abrupt risk judgments unsupported by changes in position or trajectory over time can be filtered out. 
These mechanisms help ground high-level reasoning in sensor-based evidence, thereby reducing the risk of erroneous outputs that could lead to misleading or unsafe guidance to drivers.

\section{Conclusion}
Drivers' perception of risky situations, particularly the uncertain behavior of roadside users in non-collision scenarios, remains a significant challenge that existing risk detection methods struggle to address effectively.
In this paper, we introduced VCD, which employs LLMs identify non-collision roadside risks and alerts drivers based on their eye movements.
Specifically, the system combines video processing algorithm and LLMs to identify potential risky road users on the roadside.
These risks are dynamically indicated on an adaptive HUD interface to enhance drivers' attention.
A user study with 41 drivers confirms that VCD improves drivers' risk perception and support their recognition of roadside risks.

\section*{Acknowledgments}
This paper is supported by national natural science foundation (62407038) and Aeronautical Science Fund (2023M074076001).

{\appendix[Visionary Co-Driver Prompt Details]
\label{sec:appendix_llm}
\subsection{System Prompt: Task Description}
You are Visionary Co-Driver, an expert road-scene analyst. Your goal is to help human drivers perceive pedestrian risks through natural-language reasoning.

Given a video showing multiple pedestrians and their object detection results, proceed step-by-step to:

\begin{enumerate}
    \item Identify potential risky pedestrians.
    \item Provide a binary evaluation [safe or risky] for each pedestrian ID.
\end{enumerate}

Provide your answer in Markdown format consistent with the examples provided.

\subsection{User Prompt: Input Explanation}
We have processed a driving scenario video into natural language for your reasoning. The following keys are provided for every tracked pedestrian:

\begin{itemize}
    \item ID: Globally unique identifier within the video.
    \item Surface: road, sidewalk, none (per frame range).
    \item Distance Class: Categories are very close, near, medium, far. (per frame range)  
    \item Speed Class: Categories are high, low. (per frame range)  
    \item Position Class: Categories are upper-left, upper-right, far-front, close-front, lower-left, lower-right.
\end{itemize}

\noindent Concisely describe the scene in the following format:

\begin{enumerate}
    \item Potential Risks: Describe clearly WHY any pedestrian ID poses a risk.
    \item Safety Evaluation: List each pedestrian's safety status as \texttt{Person <ID>: Safe | Risky}.
\end{enumerate}

\subsection{Example of Scene Description (JSON format)}

\texttt{Info\_roadside\_<video\_id>.json}\footnote{Generated by Grounding DINO + Segment Anything for roadside context grounding.}

\begin{Verbatim}[breaklines=true, breakanywhere=true, fontsize=\small]
{
  "road 0": { "position": ["lower-left", "close-front", "upper-left"],
              "areas pixels": 482892 },
  "road 1": { "position": ["lower-right", "close-front"],
              "areas pixels": 58185 },
  "sidewalk 0": { "position": ["lower-right", "upper-right"],
                  "areas pixels": 20968 },
  "Total objects": 17,
  "Total surface area": 562045,
  "Total person area": 31516
}
\end{Verbatim}

\texttt{person\_fusion\_<video\_id>.json}\footnote{Generated by ByteTrack + YoloX and DPT on individual pedestrians.}

\begin{Verbatim}[breaklines=true, breakanywhere=true, fontsize=\small]
[
  {
    "id": 1,
    "visible_frames": 60,
    "traj": mot_traj,
    "surface": {"1-4": "road_0", "5-40": "road_1", "41-60": "road_2"},
    "avg_depth": 8.7,
    "bbox_angle": {"1-4": 12.3, "5-40": 12.3, "81-120": 12.3},
    "distance_class": {"1-4": "near", "5-40": "far", "81-120": "near"},
    "speed_class": {"1-4": "low", "5-40": "high", "81-120": "low"},
    "position_class": {"1-4": "left up", "5-80": "middle up", "81-120": "right up"}
  },
  ...
]
\end{Verbatim}

\subsection{Few-Shot Example}

Scene Description (Urban two-lane road scenario):

\texttt{Info\_roadside\_JAAD\_video\_16.json}

\begin{Verbatim}[breaklines=true, breakanywhere=true, fontsize=\small]
{
  "road 0": { "position": ["close-front", "far-front"],
              "areas pixels": 252945 },
  "sidewalk 0": { "position": ["lower-right", "close-front"],
                  "areas pixels": 58185 },
  "sidewalk 1": { "position": ["lower-right", "upper-right"],
                  "areas pixels": 55574 },
  "Total objects": 12,
  "Total surface area": 366704,
  "Total person area": 27046
}
\end{Verbatim}

\texttt{person\_fusion\_<video\_id>.json}

\begin{Verbatim}[breaklines=true, breakanywhere=true, fontsize=\small]
[
  {
    "id": 8,
    "visible_frames": 50,
    "traj": "mot_traj_15",
    "surface": {"0-20": "sidewalk_0", "21-50": "road_0"},
    "distance_class": {"0-20": "near", "21-50": "very close"},
    "speed_class": {"0-20": "high", "21-50": "high"},
    "position_class": {"0-20": "upper-left", "21-50": "close-front"}
  },
  {
    "id": 12,
    "visible_frames": 60,
    "surface": {"0-60": "sidewalk_0"},
    "distance_class": {"0-60": "far"},
    "speed_class": {"0-60": "low"},
    "position_class": {"0-60": "upper-left"},
  }
]
\end{Verbatim}

Example Output\footnote{LLMs' raw response used for downstream HUD visualization or alert generation.}:

\begin{Verbatim}[breaklines=true, breakanywhere=true, fontsize=\small]
### Scene
Urban two-lane road; one pedestrian crosses from left sidewalk, another stays stationary on sidewalk.

#### Potential Risks
Person 8 rapidly moves from sidewalk to road (within 5m), indicating intention to cross.
Person 12 stationary on sidewalk; no immediate risk.

#### Safety Evaluation
Person 8 : Risky
Person 12 : Safe
\end{Verbatim}

\subsection{Parsed Risk JSON Format}

After LLMs reasoning, the output is structured into JSON:

\begin{Verbatim}[breaklines=true, breakanywhere=true, fontsize=\small]
{
  "video_id": "abc123",
  "interval": "0000-0014",
  "risks": [
    {"id": 8, "intention": "crossing", "risk_level": "high"},
    {"id": 12, "intention": "standing", "risk_level": "none"}
  ]
}
\end{Verbatim}
}

\bibliographystyle{IEEEtran}
\bibliography{cas-refs}

\section{Biography Section}
\begin{IEEEbiography}[{\includegraphics[width=1in,height=1.25in,clip,keepaspectratio]{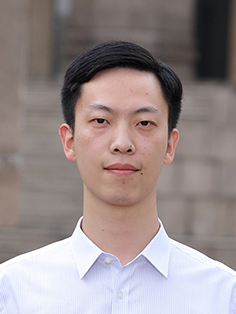}}]{Wei Xiang}
is an assistant professor in Modern Industrial Design Institute of Zhejiang University. He received his Ph.D. in digital art and design from Zhejiang University. His research interests include human AI interaction and intelligent design.
\end{IEEEbiography}
\vspace{11pt}
\begin{IEEEbiography}[{\includegraphics[width=1in,height=1.25in,clip,keepaspectratio]{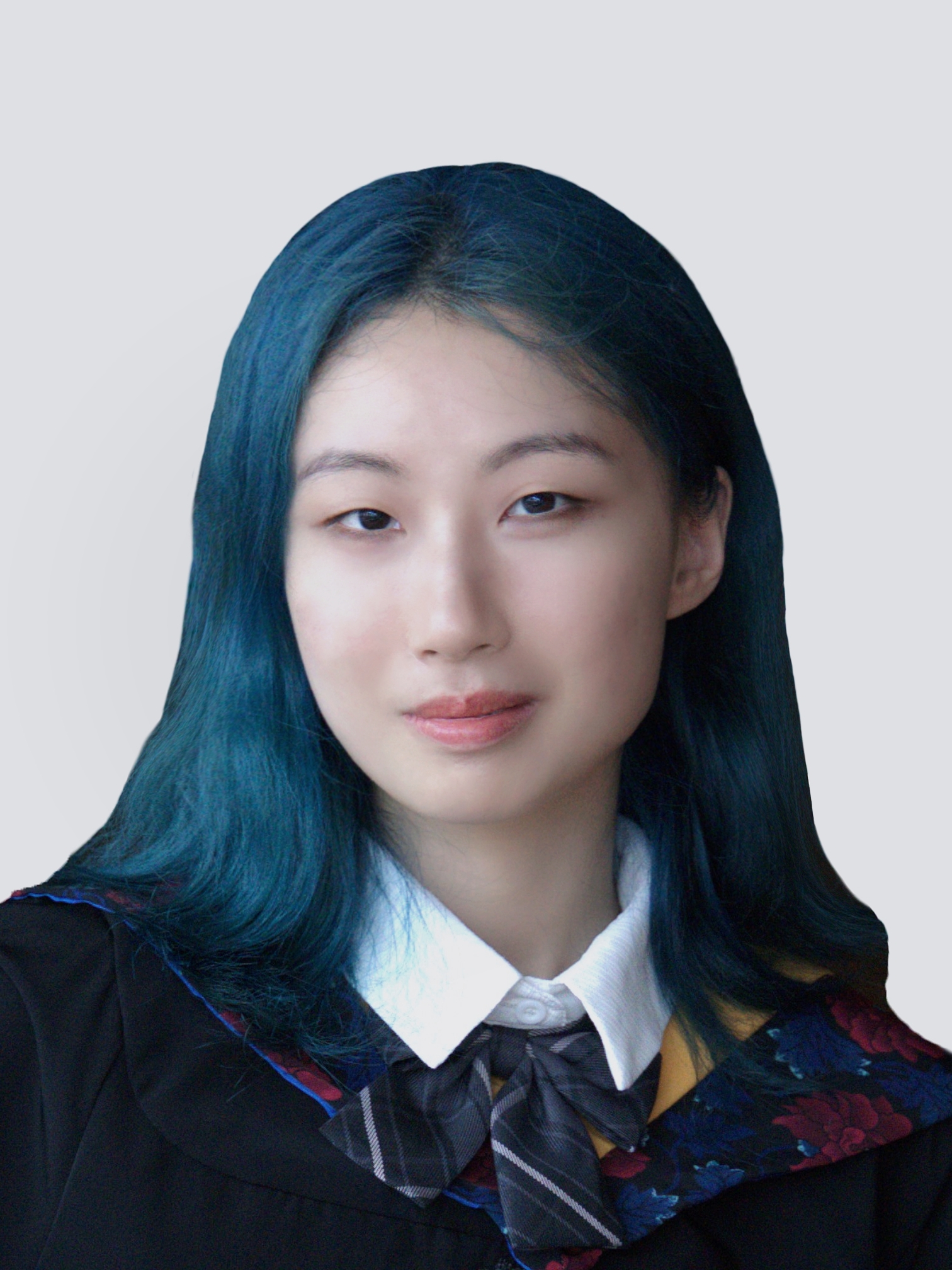}}]{Ziyue Lei}
received the Bachelor's Degree in Industrial Design from Zhejiang University in 2023, and is studying for a Master's degree in Modern Industrial Design Institute of Zhejiang University. 
Her research interests include human AI interaction and intelligent cockpit.
\end{IEEEbiography}
\vspace{11pt}
\begin{IEEEbiography}[{\includegraphics[width=1in,height=1.25in,clip,keepaspectratio]{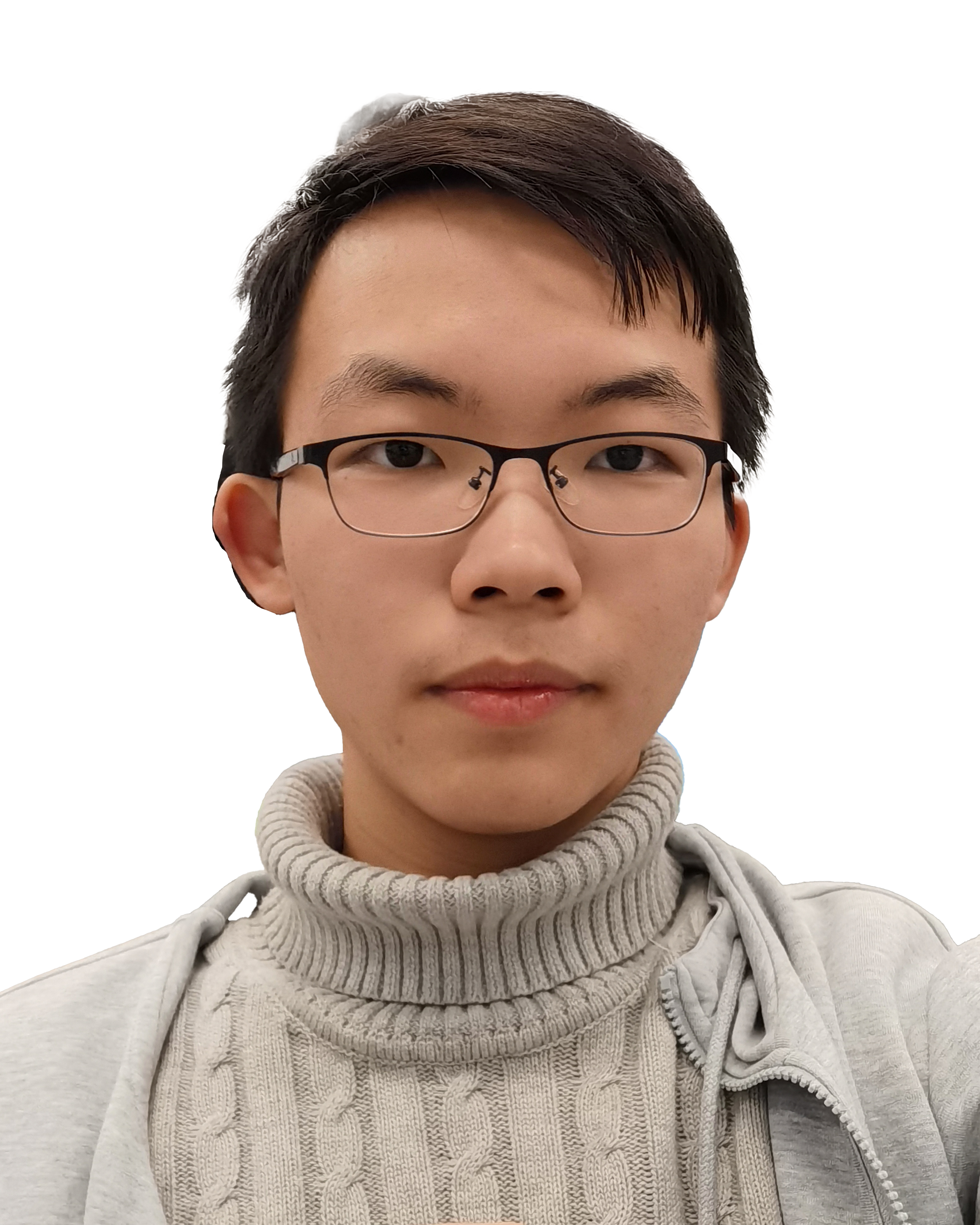}}]{Jie Wang}
 received the Bachelor of Engineering degree in Electronic and Computer Engineering from Zhejiang University, and the Bachelor of Science degree in Computer Engineering from the University of Illinois Urbana-Champaign in 2024. He is currently pursuing a Master of Science degree in Robotics at GRASP Lab, University of Pennsylvania, PA, USA.  His research interests include vision-language models, autonomous driving and robot learning. Personal website: \href{https://everloom-129.github.io/}{https://everloom-129.github.io/}
\end{IEEEbiography}
\vspace{11pt}
\begin{IEEEbiography}[{\includegraphics[width=1in,height=1.25in,clip,keepaspectratio]{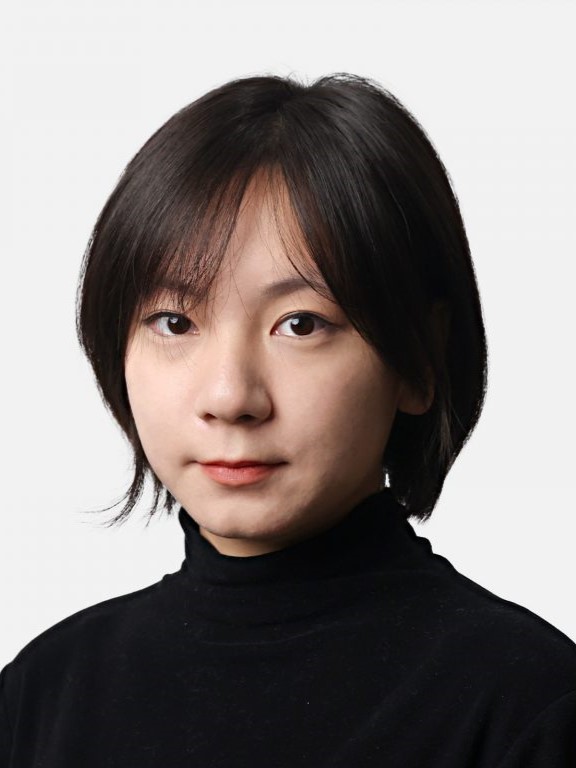}}]{Yingying Huang}
received the Bachelor of Science degree in computer science from Zhejiang University of Technology in 2021, and Master of Engineer degree in software engineering from Zhejiang University in 2024. Her research interests include human-computer interaction, natural language processing, and intelligent transportation. She is currently working at Ant Group, where she focuses on the development of large language model infrastructure and LLMs inference acceleration.
\end{IEEEbiography}
\vspace{11pt}
\begin{IEEEbiography}[{\includegraphics[width=1in,height=1.25in,clip,keepaspectratio]{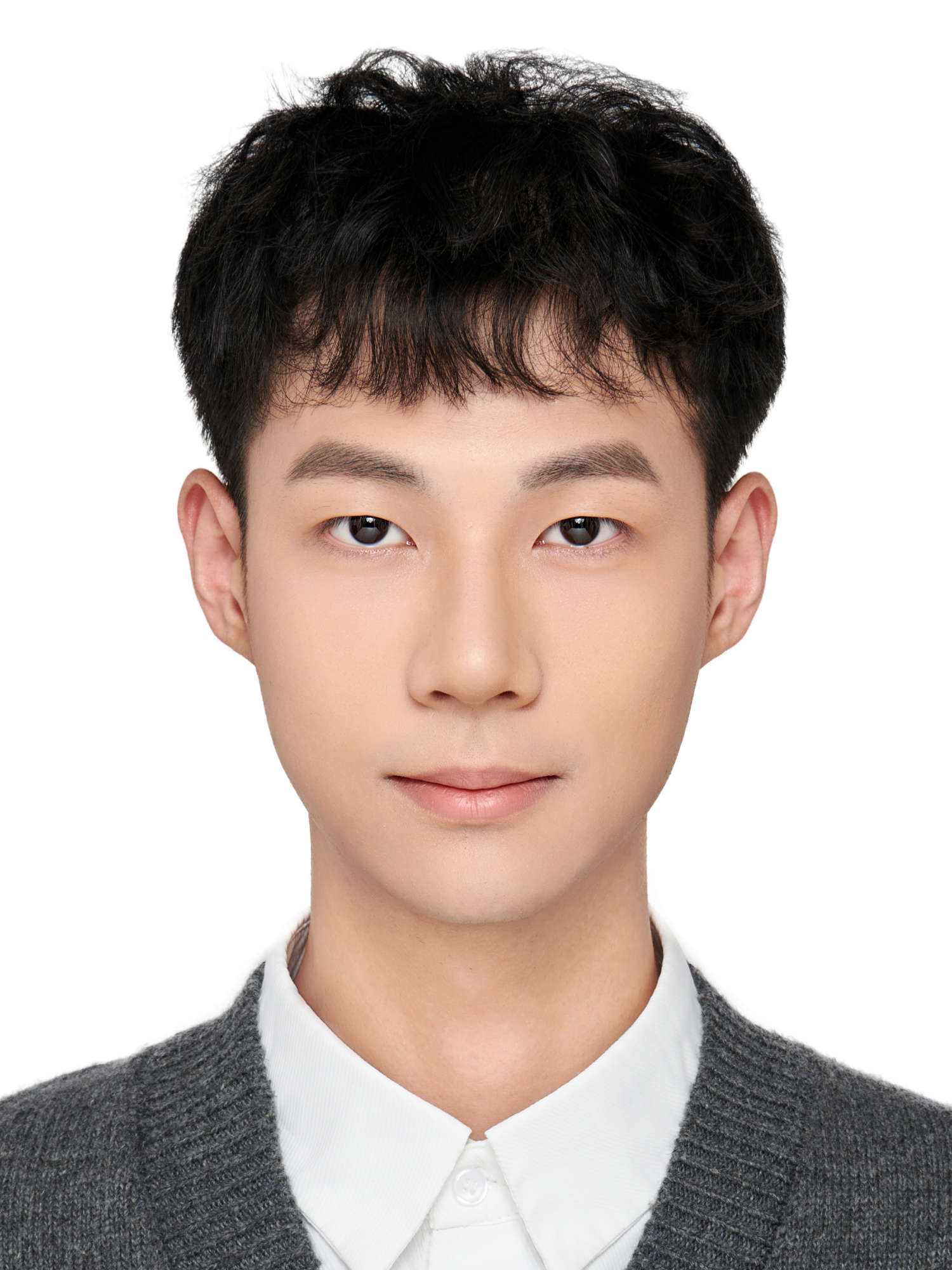}}]{Qi Zheng}
 received the Bachelor's Degree in Industrial Design from Zhejiang University, and is going to study for a Master's degree at the Hong Kong University of Science and Technology(GZ) . He is interested in HCI and AI, and would like to start research in related field. 
\end{IEEEbiography}
\vspace{11pt}
\begin{IEEEbiography}[{\includegraphics[width=1in,height=1.25in,clip,keepaspectratio]{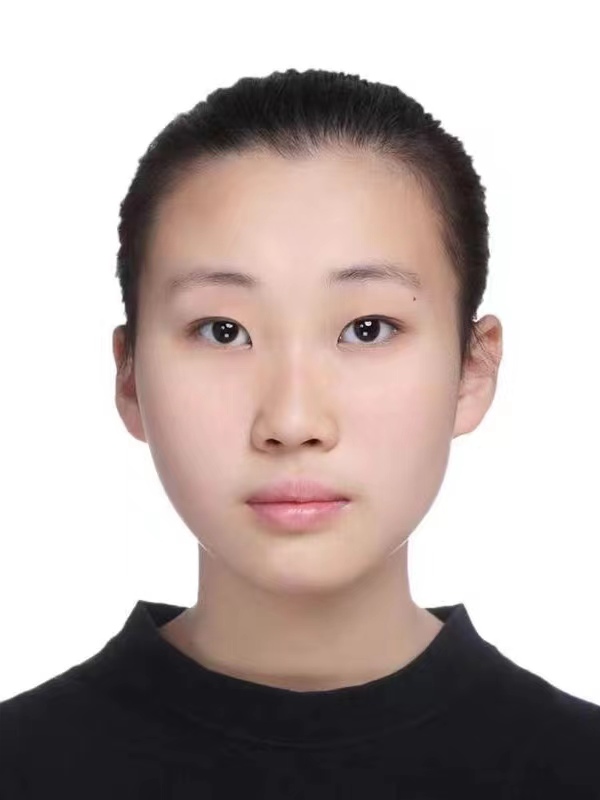}}]{Tianyi Zhang}
is currently pursuing her Bachelor of Science degree in Computer Science at Zhejiang University, China. She commenced her studies in 2022, with a focus on algorithms, and machine learning. Tianyi's research interest lies in the realm of artificial intelligence, particularly in the development of deep learning models that enhance image processing and data analysis efficiency. She aspires to continue her education with a Master's degree in Computer Science, aiming to specialize further in artificial intelligence and its applications in solving real-world problems.
\end{IEEEbiography}
\vspace{11pt}
\begin{IEEEbiography}[{\includegraphics[width=1in,height=1.25in,clip,keepaspectratio]{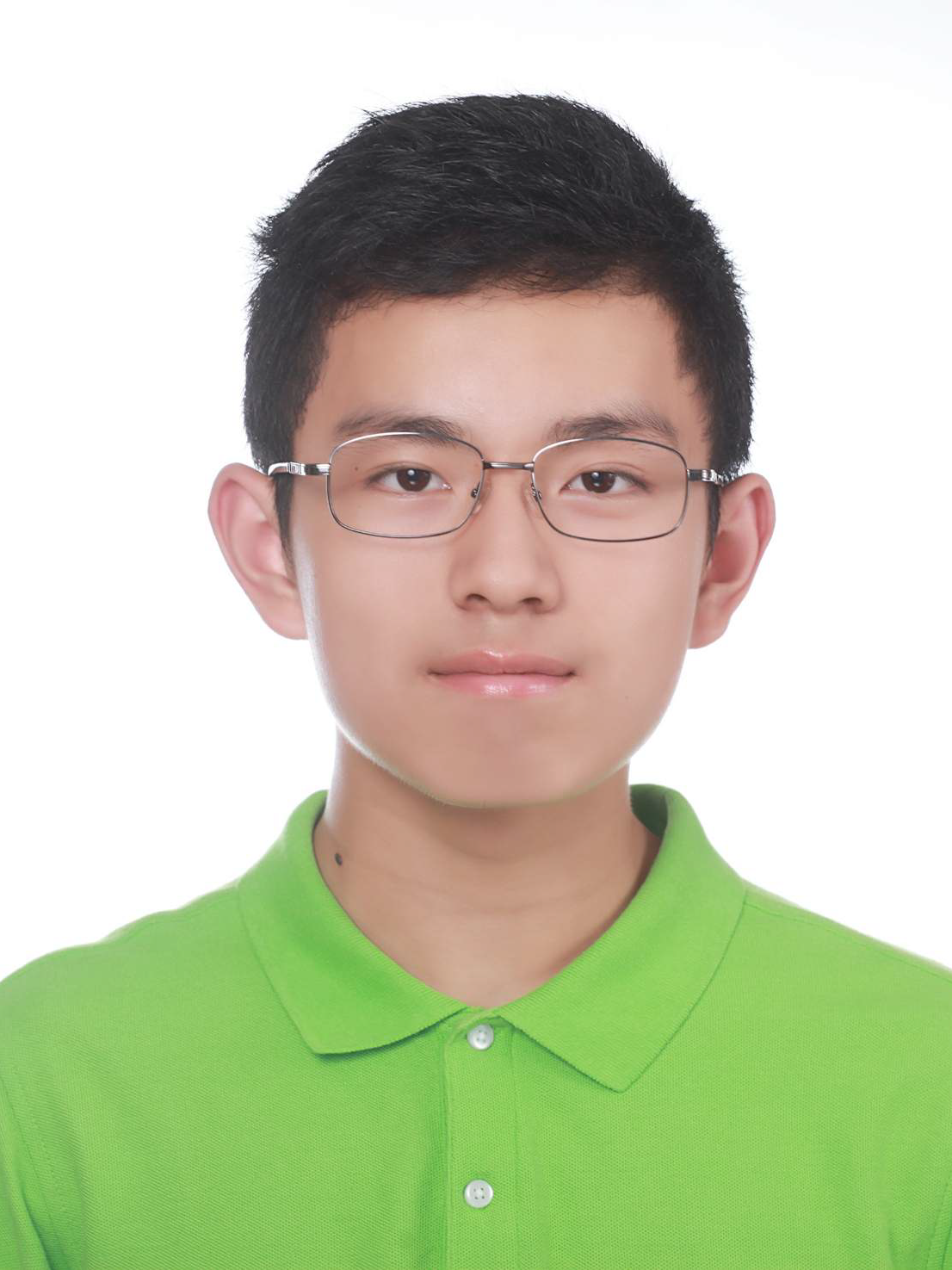}}]{An Zhao}
is currently pursuing a Bachelor of Engineering degree in Computer Science and Technology (Turing Class) at Zhejiang University, China. 
He was admitted in 2022 and is expected to graduate in 2026. 
After graduation, he will continue his academic journey as a direct doctoral student in the same institution. 
His research interests include generative artificial intelligence (AIGC), with a focus on foundation models and their applications in creative and intelligent systems.
\end{IEEEbiography}
\vspace{11pt}
\begin{IEEEbiography}[{\includegraphics[width=1in,height=1.25in,clip,keepaspectratio]{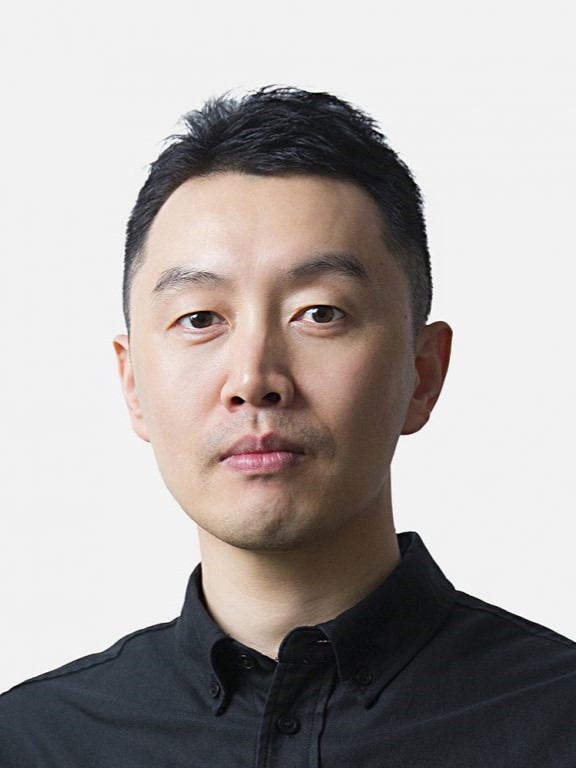}}]{Lingyun Sun}
is a Professor of Design at Zhejiang University. He’s currently a director of International Design Institute. He has interdisciplinary research experiences, including artificial intelligence, computer graphics, design cognition, interaction design, and ergonomics. He is the author or co-author of over 50 journal or conference papers on Industrial Design, Computer Aided Sketching and Interaction Design. 
\end{IEEEbiography}
\vfill

\end{document}